\documentclass{article}
\usepackage{graphicx}
\usepackage{amsmath}
\usepackage{amsthm}
\usepackage{amsfonts}
\usepackage{amssymb}

\newtheorem{thm}{Theorem}
\newtheorem{corollary}[thm]{Corollary}
\newtheorem{proposition}[thm]{Proposition}
\newtheorem{lemma}[thm]{Lemma}
\theoremstyle{definition}
\newtheorem{definition}{Definition}
\theoremstyle{remark}
\newtheorem{remark}{Remark}

\begin{document}
\baselineskip=19pt

\thispagestyle{empty}

\vspace{.5cm}

\title{Vacuum Einstein metrics with\\bidimensional Killing leaves\footnote{
Published in Diff. Geom.Appl. 16(2002)95-120 \newline Research supported in
part by the Italian Ministero dell' Universit\`{a} e della Ricerca Scientifica
e  Tecnologica.}, \\\textit{I-Local aspects}. \\  }
\author{G. Sparano $^{\text{\#,\S}}$, \ G. Vilasi $^{\text{\#,\S\S}}$, A. M.
Vinogradov $^{\text{\#,\S}}$\\{\small \ }$^{\text{\#}}${\small \
Istituto Nazionale di Fisica Nucleare, Gruppo Collegato di
Salerno, Italy.}\\{\small \ }$^{\text{\S}}${\small \ Dipartimento
di Matematica e Informatica, Universit\`{a} di Salerno, Italy.
}\\{\small \ }$^{\text{\S\S}}${\small \ Dipartimento di Scienze
Fisiche \textit{E.R.Caianiello}, Universit\`{a} di Salerno, Italy.
}\\{\small \ (e-mail: sparano@unisa.it, vilasi@sa.infn.it,
vinogradov@ponza.dia.unisa.it)}} \maketitle \vspace{1.0cm}

\noindent
\begin{abstract}
The solutions of vacuum Einstein's field equations, for the class of
Riemannian metrics admitting a non Abelian bidimensional Lie  algebra of
Killing fields, are explicitly described. They are  parametrized either by
solutions of a transcendental equation (the  {\it tortoise equation), }or
by solutions of a linear second order  differential equation in two
independent variables. Metrics,  corresponding to solutions of the tortoise
equation, are  characterized as those that admit a {\it 3-}dimensional Lie
algebra of Killing fields with bidimensional leaves.    {\it Subj. Class.:
}Differential Geometry, General Relativity.    {\it Keywords: }Einstein
metrics, Killing vectors.    {\it MSclassification}: 53C25
\vspace{5.0cm}
\end{abstract}

\vspace{8.0cm}

\vfill\eject

\section{Introduction}

In this paper we describe in an exact form local solutions (metrics) of the
vacuum Einstein equations assuming that they admit a Lie algebra
$\mathcal{G}  $ of Killing vector fields such that:
\begin{description}
\item [ I.]the distribution $\mathcal{D}$, generated by the vector fields
belonging to $\mathcal{G}$, is bidimensional,
\item[ II.] the distribution $\mathcal{D}^{\bot}$, orthogonal to $\mathcal{D}
$, is completely integrable and transversal to $\mathcal{D}$.
\end{description}

Global, in a sense, solutions of the Einstein equations constructed on the
basis of the local solutions found in this paper are discussed in the
subsequent one. There can occur two qualitatively different cases according
to  whether the dimension of $\mathcal{G}$ is $2$ or $3$. Both of them,
however,  have an important feature in common, which makes reasonable to
study them  together. Namely, all manifolds satisfying the assumptions I
and II are in a  sense fibered over $\zeta$-complex curves (see section
\ref{3G} and  \cite{SVV00}).
\begin{itemize}
\item [$\dim\mathcal{G}=2$]

Recall that, up to isomorphisms, there are two bidimensional Lie algebras:
Abelian and non-Abelian, which in what follows will be denoted by
$\mathcal{A}_{2}$ and $\mathcal{G}_{2}$ respectively.    A metric $g$
satisfying the assumptions I and II, with $\mathcal{G}
=\mathcal{A}_{2}$ or $\mathcal{G}_{2}$, will be called $\mathcal{G}
$\textit{-integrable}.   The study of $\mathcal{A}_{2}$-integrable metrics
were started by Belinsky,  Geroch, Khalatnikov, Zakharov and others
\cite{BK70}, \cite{BZ78},  \cite{Ge72}. Some remarkable properties of the
reduced, accordingly with the  above symmetry assumptions, vacuum Einstein
equations were discovered in 1978.  In particular, a suitable
generalization of the Inverse Scattering Transform,  allowed to integrate
the equations and to obtain solitary wave solutions  \cite{BZ78}. Some
physical consequences of these reduced equations were  analyzed in a number
of works (see for instance \cite{BZ79,Be91}). This paper  will be devoted
to the analysis of $\mathcal{G}_{2}$-integrable solutions, for  which some
partial results can be found in \cite{Ha88,AL92,Ch97}.    In this case, the
Killing fields ''interact'' non-trivially one another (for  instance,
$[X,Y]=Y$, for a suitable choice of the basis vectors in  $\mathcal{G}$),
while in the Abelian case these fields are absolutely free  (\textit{i.e.},
$[X,Y]=0$). Hence, it is natural to expect that the former  case is more
rigid, with respect to the latter, and, as such, it allows a more  complete
analysis. It occurs to be the case, namely, metrics in question are
parametrized by solutions of a linear equation in two independent
variables,  which, in its turn, depends linearly on a choice of a
$\zeta$-harmonic  function. Thus, this class of solutions has a ''bilinear
structure'' and,  hence, is subjected to two superposition laws.
\item[$\dim\mathcal{G}=3$]

In this case, assumption II follows automatically from I and the local
structure of this class of Einstein metrics can be explicitly described.
Some  well known exact solutions \cite{Pe69}, such as, for instance, that
of  Schwarzschild, belong to this class.
\end{itemize}

Geometrical properties of solutions described in the paper will be
discussed with more details separately.    In the paper, as it is usual,
everything is assumed to be of $C^{\infty}$  class and the following
terminological and notational convention are adopted.
\begin{itemize}
\item  manifolds are assumed to be connected and $C^{\infty}$,

\item \textit{metric }refers to a non-degenerate symmetric $\left(
0,2\right)  $ tensor field,
\item \textit{k-metric} refers to a metric on a \textit{k}-dimensional manifold,

\item  the Lie algebra of all Killing fields of a metric $g$ is denoted by
$\mathcal{K}$\textit{il}$\left(  g\right)  $ while the term \textit{Killing
algebra }refers to a subalgebra of $\mathcal{K}$\textit{il}$\left( g\right)
$,
\item \textit{integral submanifolds} of the distribution, generated by vector
fields of a Killing \ algebra $\mathcal{G}$, are called \textit{Killing
leaves,}
\item $\mathcal{A}_{2}$ stands for a bidimensional Abelian Lie algebra, while
$\mathcal{G}_{2}$ for a non-Abelian one,
\item  a $\mathcal{G}$\textit{-integrable} metric is a metric satisfying the
assumptions I and II, with $\mathcal{G}=\mathcal{A}_{2}$ or
$\mathcal{G}_{2}$.
\item  the elements of a matrix will be denoted with the corresponding lower
case letter, for instance $\mathbf{A}=\left(  a_{ij}\right)  $.
\end{itemize}

\section{Metrics admitting a bidimensional Lie algebra $\mathcal{G}_{2}$ of
Killing fields\label{ikm}.}   For a given $s\in\mathbb{R},s\neq0$, we fix a
basis $\left\{  e,\varepsilon
\right\}  $ in $\mathcal{G}_{2}$ such that $\left[  e,\varepsilon\right]
=s\varepsilon.$ It is defined uniquely up to transformations of the form
\[
e\mapsto\lambda e+\mu\varepsilon,\varepsilon\mapsto\lambda^{-1}\varepsilon
,\text{ \thinspace\thinspace\ }\lambda,\mu\in\mathbb{R},\lambda\neq0.
\]
The parameter $s$ is introduced in order to include, into our subsequent
analysis, the Abelian case ($s=0$ ) as well.    In what follows, it will be
useful the following general fact.
\begin{lemma}
Let $g$ be a metric on a differential manifold $M$. If $X\neq0$ and $fX$,
$f\in C^{\infty}\left( M\right) $, are two of its Killing fields,  then $f$
is constant. \label{lemma1}
\end{lemma}

\begin{proof}
The proof results from the formula
\begin{equation}
L_{fX}\left(  g\right)  =fL_{X}\left(  g\right)  +i_{X}\left(  g\right)
df,\label{lfx}
\end{equation} where the second term in the right hand side
is the \textit{symmetric product} of two differential \textit{1}-forms, and
$i_{X}\left(  g\right) $ the  natural insertion of $X$ in $g$. Indeed,
$L_{X}\left(  g\right)  =0$ and  $L_{fX}\left(  g\right)  =0$ imply, in
view of relation (\ref{lfx}), $i_{X}\left(  g\right)  df=0$. This shows
that $df$ vanishes at those points  where $i_{X}\left(  g\right)  \neq0$.
Since $g$ is non-degenerate, $i_{X}\left(  g\right)  $ vanishes exactly at
the same points where $X$ does.  Therefore, $df=0$, on
$\operatorname*{supp}X=\overline{\left\{ a\in M|X_{a}\neq0\right\}  }$. On
the other hand, if a Killing field vanishes on an  open subset of $M$,
then, obviously, it vanishes everywhere on $M$. For this  reason
$\operatorname*{supp}X$ is dense on $M$ and, so, $df=0$ on $M$.
\end{proof}

Let $g$ be a metric on a manifold $M$ admitting $\mathcal{G}_{2}$ as a
Killing algebra. Then, for the Killing vector fields $X$ and $Y$
corresponding,  respectively, to $e$ and $\varepsilon$, one has
\begin{equation}
\left[  X,Y\right]  =sY.\label{alg}
\end{equation}
Denote by $\mathcal{D}$ the Frobenius distribution, possibly with
singularities, generated by $X$ and $Y$.
\begin{proposition}
\label{kf}The distribution $\mathcal{D}
$ is bidimensional and in a neighborhood of a non-singular point of
$\mathcal{D}$ there exists a local chart $\left( x_{\alpha}\right  ) $ in $
M$ such that
\[
X=\partial_{n-1}\,,\quad Y=e^{sx_{n-1}}\partial_{n}.
\]
\begin{proof}
\rm{First of all, show that {\it dim}$\mathcal{D}
$ $=2$. Indeed,  in view of the above lemma if locally $X=\phi Y$, then
$\phi$  is constant and $X$ and $Y$ commute, in contradiction  with
Eq.(\ref{alg}).  Thus, the vector $Y_{a}$ and $X_{a}$ are independent for
almost all  points $a\in M$, i.e. in an everywhere dense open subset
$M_{0}$ of $M$.  Choose now a function $\phi$ such that the  fields X and
$\phi Y$ commute. In view of Eq. (\ref{alg}), this is equivalent  to
$X\left( \phi\right) +s\phi=0.$ This equation admits, obviously, a solution
in a neighborhood of any point $a\in M_{0}$.  In a local chart $\left(
y_{\mu}\right) $ in which $X=\frac{\partial}{\partial  y_{n-1}}$, $\phi
Y=\frac{\partial}{\partial y_{n}}$,  the equality $X\left(
\phi\right) +s\phi=0$ looks as $\frac{\partial\phi  }{\partial
y_{n-1}}+s\phi=0$ and hence, $\phi=e^{-sy_{n-1}+\lambda}$ where  the
function $\lambda$ does not depend on $y_{n-1}$. By passing now to
coordinates $\left( x_{\alpha}\right) $ with $x_{\alpha}=y_{\alpha}  $,
$\alpha<n$, and $x_n=\beta\left( y_{1}, s,y_{n-2},y_{n}\right  ) $ one
finds the  desired result with $\beta$ such that
$\frac{\partial\beta}{\partial y_{n} }=e^{-\lambda}$. Indeed, since
$\lambda$ does not depend on $y_{n-1} $, the last  equation admits a
solution not depending on $y_{n-1}$.}
\end{proof}
\end{proposition}
\label{prp2}

\begin{definition}
A chart of the kind introduced in the above proposition will be called
\textit{\ semi-adapted} (with respect to $X,Y$).
\end{definition}

All metrics $g$ admitting the $\left\{  X,Y\right\}  $ Killing algebra,
\textit{i.e.} such that $L_{Y}\left(  g\right)  =$ $L_{X}\left(  g\right)
=0$, are characterized by the following proposition.

\begin{proposition}
An $n$-metric $g$ admits the  vector fields $X$ and $Y$ as Killing fields
iff in a semi-adapted chart it has the following block matrix form
\[
M_{C}\left( g\right) =\left(
\begin{array}{cc}
\left( g_{ij}\right) &
\begin{array}{cc}
\;\left( sm_{i}x_{n}+l_{i}\right) & \;\;\;\;\;\;\;\left( -m_{i}\right)
\end{array}
\\
\begin{array}{c}
\left( sm_{i}x_{n}+l_{i}\right) ^{T} \\
\left( -m_{i}\right) ^{T}
\end{array}
&
\begin{array}{cc}
s^{2}\lambda x_{n}^{2}-2s\mu x_{n}+\nu& -s\lambda x_{n}+\mu\\
-s\lambda x_{n}+\mu& \lambda
\end{array}
\end{array}
\right)
\]
where $C=\{dx_{\mu}\}$ , and $g_{ij},m_{i},l_{i},\lambda,\mu,\nu,$ are
functions of $x_{l}$, $1\leq l\leq n-2$.
\end{proposition}
\label{prp3}

\begin{proof}
Indeed, the invariance with respect to $X$ shows that the components of the
metric do not depend on $x_{n-1}$ while the invariance with respect to $Y$
is  equivalent to
\begin{gather}
\partial_{n}g_{ij}=0,\,\,\,\forall i,j\leq n-2\label{inv2b0}\\
\partial_{n}g_{n-1n-1}+sg_{nn-1}=0\label{inv2b1}\\
\partial_{n}g_{n-1n}+sg_{nn}=0\label{inv2b2}\\
\partial_{n}g_{nn}=0\label{inv2b3}\\
\partial_{n}g_{in-1}+sg_{in}=0\label{inv2b4}\\
\partial_{n}g_{in}=0\label{inv2b5}
\end{gather}

Eq. (\ref{inv2b0}) tells that, for $i,j<n-1$, the components $g_{ij}$ do
not depend also on $x_{n}$, while Eqs. (\ref{inv2b1}), (\ref{inv2b2}) and
(\ref{inv2b3}), imply that, for $a,b=n-1,n$
\begin{equation}
\left(  g_{ab}\right)  =\left(
\begin{array}
[c]{cc} s^{2}\lambda x_{n}^{2}-2s\mu x_{n}+\nu & -s\lambda x_{n}+\mu\\
-s\lambda x_{n}+\mu & \lambda
\end{array}
\right)  ,
\end{equation}
where $\lambda$, $\mu$ and $\nu$ depend only on the coordinates $x_{i}$.
Eqs. (\ref{inv2b4}) and (\ref{inv2b5}) have the solution
\[
\left(
\begin{array}
[c]{cc} g_{in-1}, & g_{in}
\end{array}
\right)  =\left(
\begin{array}
[c]{cc} sm_{i}x_{n}+l_{i}\left(  x_{j}\right)  , & -m_{i}\left(
x_{j}\right)
\end{array}
\right)  .
\]
where $l_{i}^{\text{ }}$ and $m_{i}$ are arbitrary functions.
\end{proof}

For further computations it is more convenient to work with a basis, say
$\{e_{i}\}$, of vector fields invariant with respect to the Killing
algebra.  It is easy to see that all such fields are linear combinations of
\begin{equation}
e_{i}=\partial_{i},\,
\quad
e_{n-1}=\partial_{n-1}+sx_{n}\partial_{n,}
\quad
\,e_{n}=-\partial_{n}.\label{anbv}
\end{equation}
whose coefficients are $\mathcal{G}_{2}$-invariant functions, \textit{i.e.}
not depending on $x_{n-1},x_{n}$ . So, the set (\ref{anbv}) can be taken as
such a basis. Obviously, the basis of differential \textit{1}-forms
$\Theta=\left\{  \vartheta^{i}\right\}  $ dual to $\{e_{i}\}$
\begin{equation}
\vartheta^{i}=dx_{i},
\quad
\vartheta^{n-1}=dx_{n-1},
\quad
\vartheta^{n}=sx_{n}dx_{n-1}-dx_{n}.\label{anbf}
\end{equation}
is also $\mathcal{G}_{2}$-invariant. The bases (\ref{anbv}), (\ref{anbf})
are ''slightly'' non- holonomic because in the relations
\[
\left[  e_{\mu},e_{\nu}\right]  =C_{\mu\nu}^{\alpha}e_{\alpha},
\quad
d\vartheta^{\alpha}=-\frac{1}{2}C_{\mu\nu}^{\alpha}\vartheta^{\mu}
\wedge\vartheta^{\nu},
\]
all the structure constants $C_{\mu\nu}^{\alpha}$ are vanishing, except
$C_{n-1n}^{n}$, which equals $-s$. They will be called
\textit{non-holonomic  semi-adapted}.    The expression of the metric of
proposition $3$ in terms of the basis  (\ref{anbf}) is
\[
g=g_{ij}\vartheta^{i}\vartheta^{j}+\lambda\,\vartheta^{n}\vartheta^{n}
+\nu\,\vartheta^{n-1}\vartheta^{n-1}-2\mu\vartheta^{n-1}\vartheta^{n}
+2l_{i}^{\text{ }}\,\vartheta^{i}\vartheta^{n-1}+2m_{i}\vartheta^{i}
\vartheta^{n}.
\]

\begin{corollary}
An $n$-metric $g$ admits the vector fields $X$ and $Y$ as Killing fields
iff  its components, in a semi-adapted non-holonomic basis $\Theta$, do not
depend on $x_{n-1}$ and $x_{n}$. The matrix of $g$ with respect to the
basis $\Theta$ is
\begin{equation*}
\mathbf{M}_{\Theta}\left( g\right) =\left(
\begin{array}{lll}
\left( g_{ij}\right)  & \left( l_{i}^{\text{ }}\right)  & \left(
m_{i}\right)  \\
\left( l_{i}^{\text{ }}\right) ^{T} & \;\;\nu& -\mu\\
\left( m_{i}\right) ^{T} & -\mu& \;\;\lambda
\end{array}
\right) .
\end{equation*}
\end{corollary}

\section{Killing leaves\label{kl}}

The assumption II of the Introduction imposed on the metrics $g$ considered
in this paper allows, obviously, to construct semi-adapted charts, $\left\{
x_{i}\right\}  $, such that the fields $e_{i}=\frac{\partial}{\partial
x_{i}}  $, $i=1,..,n-2$, belong to $\mathcal{D}^{\bot}$. In such a chart,
called from  now on,
\textit{adapted}, the components $l_{i}$'s and $m_{i}$'s vanish. The
corresponding non-holonomic semi-adapted bases will be called
\textit{non-holonomic adapted}.

We will call \textit{orthogonal leaf }an integral (bidimensional)
submanifold of $\mathcal{D}^{\bot}$. Since $\mathcal{D}^{\bot}$ is assumed
to be  transversal to $\mathcal{D}$, the restriction of $g$ to any Killing
leaf, say  $S$, is non-degenerate. So, $\left(  S,\left.  g\right|
_{S}\right)  $ is a  homogeneous bidimensional Riemannian manifold. In
particular, the Gauss  curvature $K=K\left(  S\right)  $ of the Killing
leaves is constant. It can be  easily computed by noticing that the matrix
of the components of $\left.  g\right|  _{S}$ with respect to the chart
$\widetilde{x}=\left.  x_{n-1}\right|  _{S}$ , $\widetilde{y}=\left.
x_{n}\right|  _{S}$ is
\[
\mathbf{M}_{\left(  d\widetilde{x},d\widetilde{y}\right)  }\left(  \left.
g\right|  _{S}\right)  =\left(
\begin{array}
[c]{cc}
s^{2}\widetilde{\lambda}\widetilde{y}^{2}-2s\widetilde{\mu}\widetilde
{y}+\widetilde{\nu} & -s\widetilde{\lambda}\widetilde{y}+\widetilde{\mu}\\
-s\widetilde{\lambda}\widetilde{y}+\widetilde{\mu} & \widetilde{\lambda}
\end{array}
\right)  ,
\]
where the symbol ''\textit{tilde''} refers to the restriction to $S$ and
$\widetilde{\lambda}$, $\widetilde{\mu}$, and $\widetilde{\nu}$ are
constants  according to proposition $3$. The result is
\[
K\left(  S\right)  =\frac{\widetilde{\lambda}s^{2}}{\widetilde{\mu}
^{2}-\widetilde{\lambda}\widetilde{\nu}},\;\;\,\,\,\,\widetilde{\lambda
}\widetilde{\nu}-\widetilde{\mu}^{2}=\mathbf{M}_{\left(  d\widetilde
{x},d\widetilde{y}\right)  }\left(  \left.  g\right|  _{S}\right)  .
\]
This shows that the following cases can occur for $\left(  S,\left.
g\right|
_{S}\right)  $.

\begin{itemize}
\item [1.]$\widetilde{\lambda}>0$, $\;\widetilde{\lambda}\widetilde{\nu
}-\widetilde{\mu}^{2}>0$: $\left(  S,\left.  g\right|  _{S}\right)  $ is a
non-Euclidean plane, \textit{i.e.} a bidimensional Riemannian manifold of
negative constant Gauss curvature.
\item[2.] $\widetilde{\lambda}<0$, $\;\widetilde{\lambda}\widetilde{\nu
}-\widetilde{\mu}^{2}>0$: $\left(  S,\left.  g\right|  _{S}\right)  $ is an
''anti'' non-Euclidean plane, \textit{i.e. }is endowed with the metric of
the  previous case multiplied by $-1$.
\item[3.] $\widetilde{\lambda}\widetilde{\nu}-\widetilde{\mu}^{2}<0$: $\left(
S,\left.  g\right|  _{S}\right)  $ is any indefinite bidimensional metric
of constant Gauss curvature.
\end{itemize}

Since the Killing leaves are parametrized by $x_{1},x_{2}$, the function
\[
K=K\left(  x_{1},..,x_{n-2}\right)  =\frac{\lambda
s^{2}}{\mu^{2}-\lambda\nu}
\]
describes the behavior of the Gauss curvature when passing from one Killing
leave to another.    It is worth to note that the Killing algebra
$\mathcal{G}_{2}$ is a subalgebra  of the algebra $\mathcal{K}il\left(
g_{0}\right)  $, $g_{0}$ being a  bidimensional metric of constant
curvature (for instance, $g_{0}=\left.  g\right|  _{S}$).    If $g_{0}$ is
positive (respectively, negative) definite and of positive  (respectively,
negative) Gauss curvature, then $\mathcal{K}il\left(  g_{0}\right)  $ is
isomorphic to $so\left(  3\right)  $. But $so\left(  3\right)  $ does not
admit bidimensional subalgebras at all. This explains why  $\left. g\right|
_{S}$ cannot be a positively (respectively, negative)  curved metric in the
case ($1$) ( respectively, ($2$)).    Similarly, if $g_{0}$ is a positive
or negative definite flat metric, then  $\mathcal{K}il\left( g_{0}\right) $
admits only Abelian bidimensional  subalgebras. This explains why both
positive and negative definite flat  metrics are absent in the above list
for $\left.  g\right|  _{S}$.    In all other cases, the algebra
$\mathcal{K}il\left(  g_{0}\right)  $ admits  bidimensional non-Abelian
subalgebras.    More exactly, if $g_{0}$ is not flat, then
$\mathcal{K}$\textit{il}$\left(  g_{0}\right)  $ is isomorphic to $so\left(
2,1\right)  $. Let $\mathbf{g}$ be  the Killing form of $so\left(
2,1\right)  $. Then, the tangent planes to the  isotropic cone of
$\mathbf{g}$ exhaust the bidimensional non-Abelian Lie  subalgebras of
$so\left(  2,1\right)  $. If $g_{0}$ is flat and, thus,  indefinite, then
any bidimensional subspace of the algebra $\mathcal{K}  il\left(
g_{0}\right)  $ different from its \textit{commutator,} which is  Abelian,
is a non-Abelian subalgebra.    It is not difficult to describe the algebra
$\mathcal{K}\mathit{il}\left(
\left.  g\right|  _{S}\right)  $ in the semi-adapted coordinates $\left(
\widetilde{x},\widetilde{y}\right)  $. A direct computation shows that
$\mathcal{K}\mathit{il}\left(  g_{0}\right)  $ has the following basis:
\[
\widetilde{X}=\partial_{\widetilde{x}},\;\widetilde{Y}=e^{s\widetilde{x}
}\partial_{\widetilde{y}},\;\widetilde{Z}=e^{-s\widetilde{x}}\left[ 2\left(
s\widetilde{\lambda}\widetilde{y}-\widetilde{\mu}\right)  \partial
_{\widetilde{x}}+\left(  s^{2}\widetilde{\lambda}\widetilde{y}^{2}
-2s\widetilde{\mu}\widetilde{y}+\widetilde{\nu}\right)  \partial
_{\widetilde{y}}\right]  ,
\]
\[
\left[  \widetilde{X},\widetilde{Y}\right]  =s\widetilde{Y},\;\left[
\widetilde{X},\widetilde{Z}\right]  =-s\widetilde{Z},\;\left[  \widetilde
{Y},\widetilde{Z}\right]  =2s\widetilde{\lambda}\widetilde{X},
\]

In the case $\lambda=0$, the metric $\left.  g\right|  _{S}$ is flat
indefinite and it is convenient to identify $\left(  S,\left.  g\right|
_{S}\right)  $ with the standard plane $\left(  \mathbb{R}^{2},d\xi^{2}
-d\eta^{2}\right)  $, $\mathbb{R}^{2}=\left\{  \left(  \xi,\eta\right)
\right\}  $. To do that it is necessary to choose a bidimensional
non-commutative subalgebra in $\mathcal{K}$\textit{il}$\left(  d\xi^{2}
-d\eta^{2}\right)  $ (they are all equivalent). For instance, by choosing
$Y_{0}=\partial_{\xi}+\partial_{\eta}$, $X_{0}=-\eta\partial_{\xi}-\xi
\partial_{\eta}$, we have $\left[  X_{0},Y_{0}\right]  =Y_{0}$ , $X_{0}
,Y_{0}\in$ $\mathcal{K}$\textit{il}$\left(  d\xi^{2}-d\eta^{2}\right)  $
and, for $s\neq0$, one can identify the quadruple $\left(  S,\,2\left(
d\widetilde{x}d\widetilde{y}-\widetilde{y}d\widetilde{x}^{2}\right)
,\,\,\left.  X\right|  _{S},\,\,\left.  Y\right|  _{S}\right)  $ with
$\left(
\mathbb{R}^{2},\,\,d\xi^{2}-d\eta^{2},\,\,\,X_{0},\,\,Y_{0}\,\right)  $.

The simply connected Lie group $G$ corresponding to $\mathcal{G}$ is
isomorphic to the group of affine transformations of $\mathbb{R}$. Then,
both  $S$ and $\mathbb{R}^{2}$ are diffeomorphic to $G$ as homogeneous
$G-$spaces  and the above identification of them is an equivalence of
$G-$spaces.    The Killing form of $\mathcal{G}$ determines naturally a
symmetric covariant  tensor field on the $G-$space $G$ which is identified
with $d\widetilde{x}
^{2}$ on $S$ and with $\left(  \frac{d\xi-d\eta}{\xi-\eta}\right)  ^{2}$ on
$\mathbb{R}^{2}$. We will continue to call it \textit{Killing form}. Thus,
in the above identification the metric $\left.  g\right|  _{S}$ for
$\lambda=0$  and $s=0$ corresponds to
\begin{equation}
\widetilde{\mu}\left(  d\xi^{2}-d\eta^{2}\right)  +\widetilde{\nu}\left(
\frac{d\xi-d\eta}{\xi-\eta}\right)  ^{2}.\label{munu}
\end{equation}

This representation of the metric $\left.  g\right|  _{S}$ will be used to
describe global solutions of the Einstein equations in section
\textit{\ref{se1}.}

\section{The Ricci tensor field\label{ctr}}

In the following we will consider $4$-dimensional manifolds and will use
the following convention for the indices: Greek letters take values from
$1$ to  $4$; the first Latin letters take values from $3$ to $4$, while
$i$, $j$ from  $1$ to $2$.    Let $g$ be a $\mathcal{G}_{2}$-integrable
$4$-metric. The results of the  previous sections allow to choose a
non-holonomic adapted basis $\Theta$ such  that the matrix
$\mathbf{M}_{\Theta}\left(  g\right)  $ associated to $g$ is  of the form
\begin{equation}
\mathbf{M}_{\Theta}\left(  g\right)  =\left(
\begin{array}
[c]{cc}
\mathbf{F} & \mathbf{0}\\
\mathbf{0} & \mathbf{H}
\end{array}
\right) \label{mteta}
\end{equation}
where $\mathbf{F}$ and $\mathbf{H}$ are $2 2$ matrices whose elements
depend only on $x_{1}$ and $x_{2}$. We will distinguish two cases according
to  whether $\mathbf{F}$, \textit{i.e.,} the matrix associated to the
metric  restricted to $\mathcal{D}^{\bot}$, has negative or positive
determinant.
\begin{itemize}
\item $\det\mathbf{F}<0$. In this case, owing to the bidimensionality of
$\mathcal{D}^{\bot}$, and the independence of $\mathbf{F}$ on $x_{3}$ and
$x_{4}$, the coordinates $x_{1}$ and $x_{2}$, can be further specified to
be  characteristic coordinates on any integral submanifold of
$\mathcal{D}^{\bot}  $, so that, without changing the properties of
$\mathbf{M}_{\Theta}\left(  g\right)  $ in (\ref{mteta}), $\mathbf{F}$
takes the following form
\[
\mathbf{F}=\left(
\begin{array}
[c]{cc} 0 & f\\  f & 0
\end{array}
\right)  .
\]

\item $\det\mathbf{F}>0$. Similarly, in this case, in some isothermal
coordinates, the matrix $\mathbf{F}$ gets the form
\[
\mathbf{F}=\left(
\begin{array}
[c]{cc} f & 0\\  0 & f
\end{array}
\right)
\]
\end{itemize}

Thus, we have:
\begin{proposition}
\vspace{0in}\strut\vspace{0in}\label{gf}A $4$-metric $g,$ is $\mathcal
{G}_2$-integrable iff there exists a \textit{non-holonomic adapted basis
}$\Theta$  such that \vspace{0in}the matrix $M_{\Theta}\left( g\right  )$
of $g$ takes one of the following block  forms, according to whether
$\det{\bf F}<0$ or $\det{\bf F}>0$ .
\[
{\bf M}_{\Theta}\left( g\right) =\left(
\begin{array}{cc}
\begin{array}{cc}
0 & f \\ f & 0
\end{array}
& \mathbf{0} \\
\mathbf{0} &  {\bf H}
\end{array}
\right) ,\text{ ~}{\bf M}_{\Theta}\left( g\right) =\left(
\begin{array}{cc}
\begin{array}{cc}
f & 0 \\ 0 & f
\end{array}
& \mathbf{0} \\
\mathbf{0} & {\bf H}
\end{array}
\right)
\]
\[
{\bf H}=\left(
\begin{array}{cc}
\nu& -\mu\\
-\mu& \lambda
\end{array}
\right)
\]
$\lambda$, $\mu$, $\nu$ being arbitrary functions of $x_{i}$. In the
corresponding adapted holonomic basis $C=\left\{ dx^{\mu}\right\} $ we have
\[
{\bf M}_{C}\left( g\right) =\left(
\begin{array}{cc}
\begin{array}{cc}
0 & f \\ f & 0
\end{array}
& \mathbf{0} \\
\mathbf{0} & \overline{ {\bf H}}
\end{array}
\right) ,\text{ ~}{\bf M}_{C}\left( g\right) =\left(
\begin{array}{cc}
\begin{array}{cc}
f & 0 \\ 0 & f
\end{array}
& \mathbf{0} \\
\mathbf{0} & \overline{ {\bf H}}
\end{array}
\right)
\]
where
\[
\overline{ {\bf H}}=\left(
\begin{array}{cc}
s^{2}\lambda x^2_{4}-2s\mu x_{4}+\nu& -s\lambda x_{4}+\mu
\\
-s\lambda x_{4}+\mu& \lambda
\end{array}
\right) .
\]
\end{proposition}
\label{prp5}

It is worth to observe that $\det\overline{\mathbf{H}}=\det\mathbf{H}
=\lambda\nu-\mu^{2}$ is a functions of $x_{i}$'s only.

In the following sections the explicit expressions of the components
$R_{\mu\nu}$ of the Ricci tensor field in terms of the  function $f$ and of
the elements $h_{ab}$ of the matrix  $\mathbf{H}$ in the adapted
non-holonomic basis of corollary
\textit{5 }are found.

Recall that
\[
R_{\mu\nu}=R_{\mu\;\nu\beta}^{\;\beta}=e_{\left[  \nu\right.  }(\gamma
_{\left.  \beta\right]  \mu}^{\beta})+\gamma_{\left[  \nu\right.  \rho}
^{\beta}\gamma_{\left.  \beta\right]  \mu}^{\rho}-C_{\nu\beta}^{\rho}
\gamma_{\rho\mu}^{\beta}.
\]
with the Christoffel symbols
\begin{align*}
\gamma_{\mu\nu}^{\alpha}  & =\frac{1}{2}g^{\alpha\sigma}\left(  -e_{\sigma
}\left(  g_{\mu\nu}\right)  +e_{\mu}\left(  g_{\sigma\nu}\right)  +e_{\nu
}\left(  g_{\sigma\mu}\right)  \right) \\  & -\frac{1}{2}\left(
C_{\nu\mu}^{\alpha}+g^{\rho\alpha}g_{\sigma\mu}
C_{\nu\rho}^{\sigma}+g^{\rho\alpha}g_{\sigma\nu}C_{\mu\rho}^{\sigma}\right)
.
\end{align*}
It is easy too see that the $\gamma_{\mu\nu}^{\alpha}$'s and $R_{\mu\nu}$'s
are first order polynomials in $s$ and it is convenient to single out their
constant terms $\Gamma_{\mu\nu}^{\alpha}$ and $S_{\mu\nu},$ respectively.
More  exactly, one has:
\[
\gamma_{\mu}=\Gamma_{\mu}+\Lambda_{\mu}=\frac{1}{2}g^{-1}G_{\mu}+\Lambda_{\mu}
\]
where $\gamma_{\mu},\Gamma_{\mu},\Lambda_{\mu}$,$G_{\mu}$ are matrices
whose elements $\gamma_{\mu\nu}^{\alpha},$
$\Gamma_{\mu\nu}^{\alpha},\Lambda_{\mu
\nu}^{\alpha},G_{\mu\alpha\nu},$ are defined by

\begin{align}
\Gamma_{\mu\nu}^{\alpha}  & =\frac{1}{2}g^{\alpha\sigma}\left(  -e_{\sigma
}\left(  g_{\mu\nu}\right)  +e_{\mu}\left(  g_{\sigma\nu}\right)  +e_{\nu
}\left(  g_{\sigma\mu}\right)  \right) \nonumber\\
\Lambda_{\mu\nu}^{\alpha}  & =-\frac{1}{2}\left(  C_{\nu\mu}^{\alpha}
+g^{\rho\alpha}g_{\sigma\mu}C_{\nu\rho}^{\sigma}+g^{\rho\alpha}g_{\sigma\nu
}C_{\mu\rho}^{\sigma}\right) \label{delta}\\  G_{\mu\sigma\nu}  &
=-e_{\sigma}\left(  g_{\mu\nu}\right)  +e_{\mu}\left(  g_{\sigma\nu}\right)
+e_{\nu}\left(  g_{\sigma\nu}\right)  .\nonumber
\end{align}

\bigskip

\[
R_{\mu\nu}=S_{\mu\nu}+T_{\mu\nu}
\]

where
\begin{align*}
S_{\mu\nu}  & =e_{\left[  \nu\right.  }\Gamma_{\left.  \beta\right]  \mu
}^{\beta}+\Gamma_{\left[  \nu\right.  \rho}^{\beta}\Gamma_{\left.
\beta\right]  \mu}^{\rho}\\
T_{\mu\nu}  & =e_{\left[  \nu\right.  }\Lambda_{\left.  \beta\right]  \mu
}^{\beta}+\left(  \Gamma_{\left[  \nu\right.  }\Lambda_{\left.
\beta\right]  }\right)  _{\mu}^{\beta}+\left(  \Lambda_{\left[  \nu\right.
}\Gamma_{\left.
\beta\right]  }\right)  _{\mu}^{\beta}+\left(  \Lambda_{\left[  \nu\right.
}\Lambda_{\left.  \beta\right]  }\right)  _{\mu}^{\beta}-C_{\nu\beta}^{\rho
}\gamma_{\rho\mu}^{\beta}.
\end{align*}
Now we pass to the calculation of the Ricci tensor.
\subsection{The Ricci tensor in the case det \textbf{F }
$<$ 0}   Note that for $s=0$ the adapted non-holonomic basis becomes
holonomic and  coincides with the one used in \cite{BZ78}. This is why the
expressions for  $S_{\mu\nu}$ given below coincide with the expressions for
the components of  the Ricci tensor found in \cite{BZ78}. Observe also that
only the fields  $e_{1}$ and $e_{2}$ give nontrivial contributions to
expressions (\ref{delta})  for the $\Gamma_{\mu\nu}^{\alpha}$'s and all
components $\Lambda_{\mu\nu  }^{\alpha}$, except possibly
$\Lambda_{ba}^{c}$, vanish.
\begin{itemize}
\item \textbf{Components }$R_{ij}$:

\textbf{\ }Let us note that
\[
T_{ij}=\Lambda_{\beta\rho}^{\beta}\Gamma_{ji}^{\rho}-C_{j\beta}^{\rho}
\gamma_{\rho i}^{\beta}=0,
\]
this is due to the fact that $C_{j\beta}^{\rho}=0$, the components
$\Lambda_{\mu\nu}^{\alpha}$ with an index equal to $1$ or $2$ vanish and
$\Gamma_{ij}^{a}=0.$ So,
\[
R_{ij}=S_{ij}=e_{\left[  j\right.  }\left(  \Gamma_{\left.  \beta\right]
i}^{\beta}\right)  +\Gamma_{\left[  j\right.  \rho}^{\beta}\Gamma_{\left.
\beta\right]  i}^{\rho}.
\]
The first term of this expression gives,
\[
e_{\left[  j\right.  }\left(  \Gamma_{\left.  \beta\right]
i}^{\beta}\right)
=\partial_{j}\partial_{i}\left(  \ln\left|  f\right|  \right)  -\delta
_{ij}\partial_{i}^{2}\left(  \ln\left|  f\right|  \right)  +\partial
_{j}\partial_{i}\left(  \ln\alpha\right)  ,
\]
where $\alpha=\sqrt{\left|  \det\mathbf{H}\right|  }$ and the second term
gives
\begin{multline*}
\Gamma_{\left[  j\right.  \rho}^{\beta}\Gamma_{\left.  \beta\right]  i}^{\rho
}=tr\left(  \Gamma_{j}\Gamma_{i}\right)  -\left(  \Gamma_{\beta}\Gamma
_{j}\right)  _{i}^{\beta}=\\
\frac{1}{4}tr\left[  \mathbf{H}^{-1}\partial_{j}\left(  \mathbf{H}\right)
\mathbf{H}^{-1}\partial_{i}\left(  \mathbf{H}\right)  \right]  +\frac{\left(
\partial_{i}f\right)  ^{2}}{f^{2}}\delta_{ij}+\delta_{ij}\frac{\left(
\partial_{i}f\right)  ^{2}}{f^{2}}+\\
+\delta_{ij}\partial_{i}\left(  \ln\left|  f\right|  \right)  \partial
_{i}\left(  \ln\alpha\right)  .
\end{multline*}
Finally, one has
\begin{align*}
R_{ij}=\partial_{j}\partial_{i}\left(  \ln\left|  f\right|  \right)
-\delta_{ij}\partial_{i}^{2}\left(  \ln\left|  f\right|  \right)
+\partial_{j}\partial_{i}\left(  \ln\alpha\right)  +\\
\frac{1}{4}tr\left[  \mathbf{H}^{-1}\partial_{j}\left(  \mathbf{H}\right)
\mathbf{H}^{-1}\partial_{i}\left(  \mathbf{H}\right)  \right]  -\delta
_{ij}\partial_{i}\left(  \ln\left|  f\right|  \right)  \partial_{i}\left(
\ln\alpha\right)  .
\end{align*}

\item \textbf{Components }$R_{ab}=S_{ab}+T_{ab}$:

For what concerns $S_{ab}$, it is more convenient to use the following
expression
\[
S_{ab}=\frac{1}{\sqrt{\left|  \det g\right|  }}\partial_{\rho}\left(
\sqrt{\left|  \det g\right|  }\Gamma_{ab}^{\rho}\right)  -\partial_{a}
\partial_{b}\left(  \ln\sqrt{\left|  \det g\right|  }\right)  -\Gamma_{\rho
a}^{\beta}\Gamma_{\beta b}^{\rho}
\]
taking into account that $\left|  \det g\right|  \equiv\left|  \det
\mathbf{F}\right|  \left|  \det\mathbf{H}\right|  =f^{2}\alpha^{2}$ and
$\alpha=\sqrt{\left|  \det\mathbf{H}\right|  }$.   The result is
\[
\left(  S_{ab}\right)  =\frac{1}{2f\alpha}\mathbf{H}\left[  \left(
\alpha\mathbf{H}^{-1}\partial_{1}\left(  \mathbf{H}\right)  \right)
_{,2}+\left(  \alpha\mathbf{H}^{-1}\partial_{2}\left(  \mathbf{H}\right)
\right)  _{,1}\right]  .
\]

For $T_{ab}$ one finds
\begin{align*}
T_{ab}=e_{\left[  b\right.  }\left(  \Lambda_{\left.  \beta\right]
a}^{\beta }\right)  +\left(  \Gamma_{\left[  b\right.  }\Lambda_{\left.
\beta\right]  }\right)  _{a}^{\beta}+\left(  \Lambda_{\left[  b\right.
}\Gamma_{\left.
\beta\right]  }\right)  _{a}^{\beta}+\left(  \Lambda_{\left[  b\right.
}\Lambda_{\left.  \beta\right]  }\right)  _{a}^{\beta}-C_{b\beta}^{\rho}
\gamma_{\rho a}^{\beta}\\
=-C_{b\beta}^{\rho}\gamma_{\rho a}^{\beta},
\end{align*}
so that
\[
\left(  T_{ab}\right)  =s^{2}h_{22}\left(  \det\mathbf{H}^{-1}\right)
\mathbf{H}.
\]
and
\[
\left(  R_{ab}\right)  =\frac{1}{2f\alpha}\mathbf{H}\left[  \left(
\alpha\mathbf{H}^{-1}\partial_{2}\left(  \mathbf{H}\right)  \right)
_{,1}+\left(  \alpha\mathbf{H}^{-1}\partial_{1}\left(  \mathbf{H}\right)
\right)  _{,2}+\frac{2s^{2}}{\alpha}fh_{22}\mathbf{1}_{2}\right]  ,
\]
where $\mathbf{1}_{2}$ stands for the unit $\left(  2 2\right)  $- matrix.

\item \textbf{Components }$R_{ai}$:

\textbf{\ }In this case,
\[
S_{ai}=e_{\left[  i\right.  }\left(  \Gamma_{\left.  \beta\right] a}^{\beta
}\right)  +\Gamma_{\left[  i\right.  \rho}^{\beta}\Gamma_{\left.
\beta\right]  a}^{\rho}=0,
\]
Indeed, the first term vanishes since $\Gamma_{i}$'s are diagonal and
$\Gamma_{a}$\ are anti-diagonal. The second term also vanishes since the
matrices $\Gamma_{i}\Gamma_{j}$ are diagonal while \ $\Gamma_{i}\Gamma_{b}
$\ or \ $\Gamma_{b}\Gamma_{i}$\ anti-diagonal. Thus,
\[
R_{ai}=T_{ai}
\]
and
\begin{align*}
T_{ai}=e_{i}\left(  \Lambda_{ba}^{b}\right)  +\left(  \Gamma_{\left[
i\right.  }\Lambda_{\left.  \beta\right]  }\right)  _{a}^{\beta}+\left(
\Lambda_{\left[  i\right.  }\Gamma_{\left.  \beta\right]  }\right)
_{a}^{\beta}+\left(  \Lambda_{\left[  b\right.  }\Lambda_{\left.
\beta\right]  }\right)  _{a}^{\beta}-C_{i\beta}^{\rho}\gamma_{\rho a}^{\beta
}\\
=\left(  \Gamma_{i}\Lambda_{b}\right)  _{a}^{b}-\left(  \Lambda_{b}\Gamma
_{i}\right)  _{a}^{b}
\end{align*}
or, equivalently,
\[
\left(
\begin{array}
[c]{c} T_{3i}\\  T_{4i}
\end{array}
\right)  =s\left(
\begin{array}
[c]{c}
\left(  \mathbf{H}^{-1}\partial_{i}\left(  \mathbf{H}\right)  \right)
_{2}^{2}-\left(  \mathbf{H}^{-1}\partial_{i}\left(  \mathbf{H}\right)
\right)  _{1}^{1}\\
-2\left(  \mathbf{H}^{-1}\partial_{i}\left(  \mathbf{H}\right)  \right)
_{2}^{1}
\end{array}
\right)  .
\]
So, the final result is
\[
\left(  R_{i3},R_{i4}\right)  =s\left(  \left(  \mathbf{H}^{-1}\partial
_{i}\left(  \mathbf{H}\right)  \right)  _{2}^{2}-\left(  \mathbf{H}
^{-1}\partial_{i}\left(  \mathbf{H}\right)  \right)  _{1}^{1},\,\,-2\left(
\mathbf{H}^{-1}\partial_{i}\left(  \mathbf{H}\right)  \right)  _{2}
^{1}\right)  .
\]
\end{itemize}

The above calculations are summarized in the following proposition
\begin{proposition}
\label{ricci1}Let $g$ be a $\mathcal{G}_2$-integrable $4$-metric.
If  $\det{\bf F}<0$, then the components of the Ricci tensor in a
non-holonomic adapted basis are
\begin{eqnarray*}
\left( R_{ab}\right) &=&\frac{{\bf H}}{2f\alpha}\left[ \left(
\alpha {\bf H}^{-1}\partial_{1}\left({\bf H}\right) \right)
_{,2}+\left( \alpha {\bf H}^{-1}\partial_{2}\left( h\right)
\right) _{,1}+\frac{2s^{2}}{\alpha}
fh_{22}\mathbf{1}_{2}\right] \\ R_{12} &=&\partial_{1}\partial_{2}\left(
\ln\left| f\right| +\ln\alpha
\right) +\frac{1}{4}tr\,\left[ {\bf H}^{-1}\partial_{1}\left( {\bf H}\right)
{\bf H}^{-1}\partial_{2}\left( {\bf H}\right) \right] \\ R_{ii}
&=&-\partial_{i}\left( \ln\alpha\right) \partial_{i}\left( \ln
\left| f\right| \right) +\partial_{i}^{2}\left( \ln\alpha\right) +\frac{1
}{4}tr\,\left[ {\bf H}^{-1}\partial_{i} \left( {\bf H}\right) {\bf H}
^{-1}\partial_{i}\left(
{\bf H}\right) \right] \\
\left(
\begin{array}{l}
R_{i3} \\ R_{i4}
\end{array}
\right) &=&s\left(
\begin{array}{cc}
\left( {\bf H}^{-1}\partial_{1}\left( {\bf H}\right) \right) _{2}^{2}-\left(
{\bf H}^{-1}\partial_{1}\left( {\bf H}\right) \right) _{1}^{1} & \hspace
{0.8cm}-2\left(  {\bf H}^{-1}\partial_{1}\left( {\bf H}\right) \right)
_{2}^{1} \\
\left( {\bf H}^{-1}\partial_{2}\left( {\bf H}\right) \right) _{2}^{2}-\left(
{\bf H}^{-1}\partial_{2}\left( {\bf H}\right) \right) _{1}^{1} & \hspace
{0.8cm}-2\left(  {\bf H}^{-1}\partial_{2}\left( {\bf H}\right) \right)
_{2}^{1}
\end{array}
\right)
\end{eqnarray*}
with $\alpha=\sqrt{\left| \det{\bf H}\right| }$ .
\end{proposition}
\begin{remark}
\label{ac1}Note that for $s=0$ the above expressions for the
components of the Ricci tensor field coincide with the ones given in \cite
{BZ78}. In  particular, the components $R_{ai}$ vanish identically.
\end{remark}
\subsection{The Ricci tensor field in the case \textbf{F }
$>$0.}   We use again the adapted non-holonomic basis $\Theta$ described in
proposition
\textit{\ref{prp3},} so that the matrix of $g$ is
\[
\mathbf{M}_{\Theta}\left(  g\right)  =\left(
\begin{array}
[c]{cc}
\begin{array}
[c]{cc} 2f & 0\\  0 & 2f
\end{array}
& \mathbf{0}\\
\mathbf{0} & \mathbf{H}
\end{array}
\right)  =\left(
\begin{array}
[c]{cc}
\mathbf{F} & \mathbf{0}\\
\mathbf{0} & \mathbf{H}
\end{array}
\right)  .
\]

In this case essentially the same computation as before gives the following
result.
\begin{proposition}
\label{ricci2}Let $g$ be a $\mathcal{G_2}$-integrable $4$-metric. If
$\det{\bf F}
>0$, then the components of the Ricci tensor in a non-holonomic adapted
basis are
\begin{align*}
\left( R_{ia}\right) & =s\left(
\begin{array}{cc}
\left({\bf H}^{-1}\partial_{1}\left({\bf H}\right) \right) _{2}^{2}-\left(
{\bf H}^{-1}\partial_{1}\left( {\bf H}\right) \right) _{1}^{1} & -2\left(
{\bf H}^{-1}\partial_{1}\left( {\bf H}\right) \right) _{2}^{1} \\
\left( {\bf H}^{-1}\partial_{2}\left({\bf H}\right) \right) _{2}^{2}-\left(
{\bf H}^{-1}\partial_{2}\left({\bf H}\right) \right) _{1}^{1} & -2\left(
{\bf H}^{-1}\partial_{2}\left({\bf H}\right) \right) _{2}^{1}
\end{array}
\right) ; \\
\left( R_{ab}\right) & =\frac{{\bf H}}{2f\alpha}\left[ \frac{1}{2}\left
[ \left(
\alpha{\bf H}^{-1}\partial_{1}\left( {\bf H}\right) \right) _{,1}
+\left( \alpha {\bf H}^{-1}\partial_{2}\left({\bf H}h\right) \right)
_{,2}\right  ] +\frac{2s^{2}}{\alpha}fh_{22}\mathbf{1}_{2}\right] ; \\
R_{11}& =\frac{1}{2}\left[ \triangle\left( \ln\alpha\ln\left| f\right|
\right) +\frac{1}{2}tr\,\left( {\bf H}^{-1}\partial_{1} {\bf H}\right
) ^{2}-\frac{\alpha ,_{1}}{\alpha}\partial_{1}\left( \ln\left| f\right|
\right) \right] + \\  & +\frac{1}{2}\left[
\frac{\alpha,_{2}}{\alpha}\partial_{2}\left( \ln
\left| f\right| \right) +\partial_{1}\left( \frac{\alpha,_{1}}{\alpha}
\right) -\partial_{2}\left( \frac{\alpha,_{2}}{\alpha}\right) \right] ; \\
R_{22}& =\frac{1}{2}\left[ \triangle\left( \ln\alpha\ln\left| f\right|
\right) +\frac{1}{2}tr\left( \,{\bf H}^{-1}\partial_{2}{\bf H}\right
) ^{2}+\frac{\alpha ,_{1}}{\alpha}\partial_{1}\left( \ln\left| f\right|
\right) \right] + \\  & -\frac{1}{2}\left[
\frac{\alpha,_{2}}{\alpha}\partial_{2}\left( \ln
\left| f\right| \right) -\partial_{1}\left( \frac{\alpha,_{1}}{\alpha}
\right) +\partial_{2}\left( \frac{\alpha,_{2}}{\alpha}\right) \right] ; \\
R_{12}& =\frac{1}{2}\left[ -\frac{\alpha,_{1}}{\alpha}\partial_{2}\left(
\ln\left| f\right| \right) -\frac{\alpha,_{2}}{\alpha}\partial_{1}\left(
\ln\left| f\right| \right) +2\partial_{1}\partial_{2}\left( \ln\alpha
\right) \right] + \\
& +\frac{1}{4}tr\left[ \,{\bf H}^{-1}\partial_{1}\left({\bf H}\right){\bf
H}^{-1}\partial
_{2}\left({\bf H}\right) \right] ;
\end{align*}
with
\[
\triangle= \frac{\partial^{2}}{\partial x_{1}^{2}}+\frac{\partial
^{2}}{\partial x_{2}^{2}} .
\]
\end{proposition}

\begin{remark}
\label{ac2}Also in this case the components $R_{ai}
$ vanish identically for  $s=0$.
\end{remark}

\section{Solutions of vacuum Einstein field equations\label{se1}}

In this section we will limit ourselves to discuss only the general form of
local solutions of vacuum Einstein equations
\[
R_{\mu\nu}=0
\]
for $\mathcal{G}_{2}$-integrable \textit{normal} (see after) metrics.   Let
us consider separately the cases characterized by $\det\mathbf{F}<0$ and
$\det\mathbf{F}>0$ .
\subsection{Solutions of Einstein equations in the case det \textbf{F }
$<$\textbf{0}}

Note that, for $s=0$ (Abelian case) the equations $R_{ai}=0$ become
identities, while for $s\neq0$ they impose the following strong conditions
on  the metric:
\begin{equation}
\left\{
\begin{array}
[c]{l}
\left(  \mathbf{H}^{-1}\partial_{i}\left(  \mathbf{H}\right)  \right)
_{2}^{2}=\left(  \mathbf{H}^{-1}\partial_{i}\left(  \mathbf{H}\right)
\right)  _{1}^{1}\\
\left(  \mathbf{H}^{-1}\partial_{i}\left(  \mathbf{H}\right)  \right)
_{2}^{1}=0
\end{array}
\right.  .\label{rai}
\end{equation}
The two cases $h_{22}\neq0$ and $h_{22}=0$ are qualitatively different and
will be discussed separately.
\subsubsection{The case $h_{22}\neq0\label{se1.1}$}

In this case equations (\ref{rai}) imply that $\left(  \mathbf{H}^{-1}
\partial_{i}\left(  \mathbf{H}\right)  \right)  _{1}^{2}=0$ for any symmetric
$\left(  2 2\right)  $-matrix $\mathbf{H}$. This means that
$\mathbf{H}^{-1}\partial_{i}\left(  \mathbf{H}\right)  $ is a scalar
matrix,
\textit{i.e.,}
\[
\partial_{1}\left(  \mathbf{H}\right)  =\varphi\mathbf{H,\,\,\,\,\,\,}
\partial_{2}\left(  \mathbf{H}\right)  =\psi\mathbf{H}
\]
for some functions $\varphi=\varphi\left(  x_{i}\right)  $,
$\psi=\psi\left( x_{i}\right)  $.    The compatibility condition
$\partial_{2}\left(  \varphi\right)  =\partial
_{1}\left(  \psi\right)  $ for the above system, implies the existence
(locally) of a function $\gamma\left(  x_{i}\right)  $ such that
$\varphi=\partial_{1}\left(  \gamma\right)  $, $\psi=\partial_{2}\left(
\gamma\right)  $ . The function $\gamma$ can be chosen in such a way that
$\mathbf{H}=e_{\,}^{\gamma}\mathbf{M}$, $\mathbf{M}$ being a constant
symmetric $\left(  2 2\right)  $-matrix such that $\det\mathbf{M}=\pm1$.
Thus,
\[
\alpha=e^{\gamma}.
\]
Then the equations $R_{ab}=0$ can be written as
\begin{equation}
\alpha,_{12}+s^{2}fm_{22}=0,\label{na1}
\end{equation}
or
\[
f=c\alpha,_{12},
\]
$\alpha,_{i}\equiv\partial_{i}\left(  \alpha\right)  $ , $\alpha,_{ij}
\equiv\partial_{i}\partial_{j}\left(  \alpha\right)  $, and
\[
c=-\frac{1}{s^{2}m_{22}}.
\]

This brings Einstein equations to the form
\begin{align}
\mathbf{H}  & =e^{\gamma}M=\alpha\mathbf{M}\label{ce1}\\
f  & =c\alpha,_{12}\label{ce1,1}\\
\partial_{i}\left(  \ln\left|  f\right|  \right)   & =\partial_{i}\left(
\ln\frac{\left|  \alpha,_{i}\right|  }{\sqrt{\alpha}}\right) \label{ce2}\\
\partial_{1}\partial_{2}\left(  \ln\left|  f\right|  \right)   & =-\frac
{1}{\alpha}\alpha,_{12}+\frac{1}{2\alpha^{2}}\alpha,_{1}\alpha,_{2}
.\label{ce3}
\end{align}

For the two possible values of the index $i$ Eq. (\ref{ce2}) gives
\begin{align}
f  & =H\left(  x_{2}\right)  \partial_{1}\left(  \sqrt{\alpha}\right)
\label{h}\\
& =K\left(  x_{1}\right)  \partial_{2}\left(  \sqrt{\alpha}\right)
\label{k}
\end{align}
where $H$ and $K$ are arbitrary functions, or, equivalently,
\begin{equation}
H\partial_{1}\alpha=K\partial_{2}\alpha.\label{haka}
\end{equation}
From Eq. (\ref{ce1,1}) one gets
\[
\alpha,_{1}=\frac{1}{c}K\left(  \sqrt{\alpha}-A\right)  ,\qquad\alpha
,_{2}=\frac{1}{c}H\left(  \sqrt{\alpha}-A\right)  ,
\]
where $A$ is a constant, or, equivalently,
\[
d\alpha=\frac{1}{c}\left(  \sqrt{\alpha}-A\right)  \left(  Kdx_{1}
+Hdx_{2}\right)  .
\]
By setting $\beta^{2}=\alpha$ the above equation integrates to the equality

\[
\beta+A\ln\left|  \beta-A\right|  =F\left(  x_{1}\right)  +G\left(
x_{2}\right)  ,
\]
with $F\left(  x_{1}\right)  \equiv\frac{1}{2c}\int Kdx_{1},
\quad
G\left(  x_{2}\right)  \equiv\frac{1}{2c}\int Hdx_{2}.$ The above equation
will be called the \textit{tortoise equation}. Finally, the remaining
Einstein  equations show Eq. (\ref{ce3}) to be an identity.    By summing
up we give the components of the metric in the basis $C=\left\{
dz_{1},dz_{2},dx,dy\right\}  $ with $z_{1}=\frac{1}{2}\left(  x_{1}
+x_{2}\right)  ,$ $z_{2}=\frac{1}{2}\left(  x_{1}-x_{2}\right)  ,$
$x=x_{3},$  $y=x_{4},$ where the $x_{\mu}$'s are the adapted coordinates
mentioned in  proposition \textit{\ref{prp5}}.
\begin{proposition}
\label{gs1}Any $\mathcal{G}
_2$-integrable $4$-metric $g$ satisfying the vacuum Einstein equations,
and such that $\det{\bf F}<0$ and $h_{22}\neq0$, has in the adapted
coordinate $\left(z_1,z_2,x,y\right  )$ the following matrix form
\[
{\bf M}_{C}\left( g\right) =\left(
\begin{array}{cc}
\begin{array}{cc}
2f & 0 \\ 0 & -2f
\end{array}
& \mathbf{0} \\
\mathbf{0} & \beta^2 \left(
\begin{array}{cc}
s^{2}ky^{2}-2sly+m & -sky+l \\
-sky+l & k
\end{array}
\right)
\end{array}
\right)
\]
where
\begin{itemize}
\item$k$, $l$, $m$, are arbitrary constants such that $km-l^{2}=\pm1$,
$k\neq0$,
\item
\begin{equation}
f=-\frac{1}{4s^{2}k}\left( \frac{\partial^{2}}{\partial z_1^{2}}-\frac{
\partial^{2}}{\partial z_2^{2}}\right) \beta^2  ,  \label{sp0}
\end{equation}
\item$\beta$ is a solution of the tortoise equation
\begin{equation}
\beta+A\ln\left|\beta-A\right| =F\left( z_1+z_2\right)
+G\left( z_1-z_2\right) ,  \label{sp1}
\end{equation}
$A$,  $F$, $G$ being an arbitrary constant and arbitrary functions
respectively.
\end{itemize}
\end{proposition}

\begin{remark}
As it will be clarified  in \cite{SVV00}, the {\it tortoise equation}
(\ref{sp1}) leads to a  deeper understanding of the so called Regge-Wheeler
tortoise coordinate,  which, apart from constant terms, is defined as its
left hand side.
\end{remark}

\begin{remark}
Concerning the signature of the metric and the character of the Killing
fields, we observe that:  If $\det\mathbf{M}=1$ (see Eq. (\ref{ce1})), then
$\mathbf{H}$ is either  positive or negative definite according to the sign
of $k$ and  $g\left( Y,Y\right) $, $g\left( X,X\right) $ have the same sign
as $k$. The signature of $g$ is equal to $\pm2$, so that these metrics are
of interest for general relativity;  If $\det\mathbf{M}=-1$, then
$\mathbf{H}$ is indefinite, $g\left(  Y,Y\right) $ has again the same sign
as $k$ while the sign of $g\left(  X,X\right) $ varies depending on  the
values of $y$. The signature of $g$ in this case is equal to $0$.
\end{remark}

By using the results of section \ref{kl}, we have:
\begin{corollary}
The metric $g$ of the above proposition admits an additional Killing field
\begin{equation*}
Z=e^{-sx}\left[ 2\left( sky-l\right) \partial_{x}+\left(
s^{2}ky^{2}-2sly+m\right) \partial_{y}\right] ,
\end{equation*}
which generates together with $X=\partial_{x}$ and $Y=e^{sx}\partial_{y}$ a
3\textit{-dimensional Lie algebra} isomorphic to $so\left(2,1\right  )$
(assuming that $s\neq0$):
\begin{equation*}
\left[ X,Y\right] =sY,\,\,\,\left[ X,Z\right] =-sZ,\,\,\left[ Y,Z\right]
=2skX
\end{equation*}
\end{corollary}

\subsubsection{The case $h_{22}=0$.\label{se1.2}}

Now, Eqs. (\ref{rai}) are identically satisfied, while the remaining
Einstein equations become
\begin{gather}
\left(  \alpha\mathbf{H}^{-1}\partial_{1}\mathbf{H}\right)  ,_{2}+\left(
\alpha\mathbf{H}^{-1}\partial_{2}\mathbf{H}\right)  ,_{1}=0\label{bz}\\
\partial_{1}\partial_{2}\left(  \ln\left|  f\right|  +\ln\alpha\right)
+\frac{1}{4}tr\,\left[  \mathbf{H}^{-1}\partial_{1}\left(
\mathbf{H}\right)
\mathbf{H}^{-1}\partial_{2}\left(  \mathbf{H}\right)  \right]  =0\label{ce}\\
-\partial_{i}\ln\left|  \alpha\right|  \partial_{i}\ln\left|  f\right|
+\partial_{i}^{2}\ln\alpha+\frac{1}{4}tr\left[  \,\mathbf{H}^{-1}\partial
_{i}\left(  \mathbf{H}\right)  \mathbf{H}^{-1}\partial_{i}\left(
\mathbf{H}\right)  \right]  =0.\label{fe}
\end{gather}
In terms of the components $\mu$ and $\nu$ of $\mathbf{H}$ they reduce to
\begin{subequations}
\begin{gather}
\alpha,_{12}=0\label{h01}\\
\left(  \alpha w_{,1}\right)  _{,2}+\left(  \alpha w_{,2}\right)
_{,1}=0\label{h02}\\
\partial_{1}\partial_{2}\left(  \ln\left|  f\right|  \right)  =\frac
{\alpha,_{2}\alpha,_{1}}{2\alpha^{2}}\label{h03}\\
\alpha,_{i}\partial_{i}\left(  \ln\left|  f\right|  \right)  =\alpha
,_{ii}-\frac{\alpha,_{i}^{2}}{2\alpha},\label{h04}
\end{gather}
with $\alpha=\sqrt{\left|  \det\mathbf{H}\right|  }=\left|  \mu\right|  $
and $w=\frac{\nu}{\alpha}$.    The general solution of Eq. (\ref{h01}) is
\end{subequations}
\[
\alpha=F\left(  x_{1}\right)  +G\left(  x_{2}\right)  ,
\]
$F$ and $G$ being arbitrary functions such that $\alpha$ is positive.   The
general solution of Eq. (\ref{h03}) is
\[
f=\pm\alpha^{-\frac{1}{2}}e^{P\left(  x_{1}\right)  +Q\left(  x_{2}\right)
}
\]
where $P$ and $Q$ are arbitrary functions.   Now equation (\ref{h04}) takes
the form
\begin{align*}
P^{\prime}\left(  x_{1}\right)  \alpha,_{1}  & =\alpha,_{11}\\
Q^{\prime}\left(  x_{2}\right)  \alpha,_{2}  & =\alpha,_{22}
\end{align*}
and are resolved as
\[
F=C_{1}\int e^{P}dx_{1}+D_{1}
\quad
G=C_{2}\int e^{Q}dx_{2}+D_{2}.
\]
Thus as the final result we see that the general solution of the
differential system (\ref{h01}), (\ref{h03}), (\ref{h04}) is given by
\begin{align*}
\alpha & =C_{1}\int e^{P}dx_{1}+C_{2}\int e^{Q}dx_{2}+C\\
f  & =\pm\alpha^{-\frac{1}{2}}e^{P\left(  x_{1}\right)  +Q\left(
x_{2}\right)  }
\end{align*}
where $C$, $C_{1}$, $C_{2}$, are arbitrary constants such that $\alpha$ is
positive.   Eq. (\ref{h02}) is a \textit{linear second order partial
differential  equation} and can be studied by standard methods. We postpone
this problem to  a further publication.    As in proposition
\textit{\ref{gs1}} we summarize the obtained results by  giving the
components of $g$ in the frame $C=\left\{  dz_{1},dz_{2}  ,dx,dy\right\} $
where $z_{1}=\frac{1}{2}\left(  x_{1}+x_{2}\right)  ,$
$z_{2}=\frac{1}{2}\left(  x_{1}-x_{2}\right)  ,$ $x=x_{3},$ $y=x_{4},$ and
$x_{\mu}$' s are the adapted coordinates introduced in proposition
\ref{prp5}.
\begin{proposition}
\label{gs2}Any $\mathcal{G}
_2$-integrable $4$-metric $g$ satisfying the vacuum
Einstein equations and such that $\det{\bf F}<0$ and $h_{22}=0$, has the
following matrix form in the  adapted coordinates $\left(
z_1,z_2,x,y\right) $,
\[
{\bf M}_{C}\left( g\right) =\left(
\begin{array}{cc}
\begin{array}{cc}
2f & 0 \\ 0 & -2f
\end{array}
& \mathbf{0} \\
\mathbf{0} & \mu\left(
\begin{array}{cc}
-2sy+w & 1 \\
1 & 0
\end{array}
\right)
\end{array}
\right)
\]
where
\begin{itemize}
\item
\begin{eqnarray}
\mu&=&C_{1}F\left( z_1+z_2\right) +C_{2}G\left( z_1-z_2\right) +C  \label{mf1}
\\
f &=&\left| \mu\right| ^{-\frac{1}{2}}F^{\prime}G^{\prime}  \label{mf2}
\end{eqnarray}
$F$, $G$ and $C$, $C_{1}$, $C_{2}$ being arbitrary functions and arbitrary
constants respectively, such that $\mu  $ and $f$  are everywhere
nonvanishing;
\item$w$ is an arbitrary solution of the equation
\[
\mu\left( \frac{\partial^{2}}{\partial z_1^{2}}-\frac{\partial^{2}}{
\partial z_2^{2}}\right) w+\frac{\partial\mu}{\partial z_1}\frac{\partial
w}{
\partial z_1}-\frac{\partial\mu}{\partial z_2}\frac{\partial w}{\partial
z_2}=0.
\]
\end{itemize}
\end{proposition}

In this case, $\det\mathbf{H}<0$ and the metric $g$ has signature equal to
$0$. The Killing field $Y$ is isotropic, while the sign of $g\left(
X,X\right)  $ varies as a function of $y$. The curvature $K$ of the Killing
leaves vanishes.
\begin{remark}
In contrast with the case $h_{22}\neq 0$ (see  5.1.1) an additional Killing
field, say $Z$,  tangent to the Killing leaves and independent  on $X$ and
$Y$ exists only if $w$ is a  constant, say $w_0$. In  such a case
\begin{equation*}
Z=e^{-sx}\left[ -2 \partial_{x}+\left(
-2sy+w_0\right) \partial_{y}\right] ,
\end{equation*}
and generates together to $X=\partial_{x}$ and $Y=e^{sx}\partial_{y}$ a
3\textit{-dimensional Lie algebra} isomorphic to $\mathcal{K}il \left
(dx^2-dy^2\right)$:
\begin{equation*}
\left[ X,Y\right] =sY,\,\,\,\left[ X,Z\right] =-sZ,\,\,\left[ Y,Z\right]
=0
\end{equation*}
\end{remark}

A canonical form for Eq. (\ref{h02}) may be obtained by passing to
coordinates
\[
\xi=F\left(  x_{1}\right)  ,\hspace{0.2in}\eta=G\left(  x_{2}\right)
\]
in which Eq. (\ref{h02}) becomes
\[
2\left(  \xi+\eta\right)  \frac{\partial^{2}\widetilde{w}}{\partial\xi
\partial\eta}+\frac{\partial\widetilde{w}}{\partial\xi}+\frac{\partial
\widetilde{w}}{\partial\eta}=0,
\]
with $\widetilde{w}\left(  \xi,\eta\right)  \equiv w\left(  F^{-1}\left(
\xi\right)  ,G^{-1}\left(  \eta\right)  \right)  $, or, alternatively,
\[
\frac{\partial^{2}Z}{\partial\xi\partial\eta}+\frac{1}{4\left(  \xi
+\eta\right)  ^{2}}Z=0,\qquad Z=\sqrt{\xi+\eta}\widetilde{w}.
\]

Its geometrical interpretation is given in \cite{SVV00}.
\subsection{Solutions of Einstein equations in the case det \textbf{F
$>$ 0}}   As before, the equations $R_{ai}=0$ are satisfied trivially if
$s=0$ while for  $s\neq0$ they coincide with (\ref{rai}):
\begin{equation}
\left\{
\begin{array}
[c]{l}
\left(  \mathbf{H}^{-1}\partial_{i}\left(  \mathbf{H}\right)  \right)
_{2}^{2}=\left(  \mathbf{H}^{-1}\partial_{i}\left(  \mathbf{H}\right)
\right)  _{1}^{1}\\
\left(  \mathbf{H}^{-1}\partial_{i}\left(  \mathbf{H}\right)  \right)
_{3}^{2}=0
\end{array}
\right.  .\label{rai2}
\end{equation}
Again it is convenient to treat separately the cases $h_{22}\neq0$ and
$h_{22}=0$.
\subsubsection{The case $h_{22}\neq0$}

As in sec. \ref{se1.1}, equations $R_{ia}=0$ are solved as
\[
\mathbf{H}=e^{\gamma}\mathbf{M}_{\,}.
\]
$\mathbf{M}$ being a constant symmetric $\left(  2 2\right)  $-matrix  such
that $\det M=\pm1$ and $\alpha=e^{\gamma}.$ Because of the non-degeneracy
of $g$ the first derivatives of $\alpha$ are non-vanishing, so that
Einstein  equations can be brought to the following form
\begin{gather}
\mathbf{H}=\alpha\mathbf{M}_{\,},\label{aee1}\\
\left(  \frac{\triangle\left(  \alpha\right)  }{4f}+s^{2}m_{22}\right)
\mathbf{M=0},\label{aee2}\\
\triangle\left(  \ln\alpha\left|  f\right|  \right)  -\frac{1}{\alpha
f}\left(  \alpha,_{1}f,_{1}-\alpha,_{2}f,_{2}\right)  +\frac{\left(
\alpha,_{2}\right)  ^{2}}{\alpha^{2}}+\frac{\alpha,_{11}-\alpha,_{22}}{\alpha
}=0,\label{aee3}\\
\triangle\left(  \ln\alpha\left|  f\right|  \right)  +\frac{1}{\alpha
f}\left(  \alpha,_{1}f,_{1}-\alpha,_{2}f,_{2}\right)  +\frac{\left(
\alpha,_{1}\right)  ^{2}}{\alpha^{2}}-\frac{\alpha,_{11}-\alpha,_{22}}{\alpha
}=0,\label{aee4}\\
\frac{1}{2\alpha f}\left(  \alpha,_{1}f,_{2}+\alpha,_{2}f,_{1}\right)
+\frac{\alpha,_{2}\alpha,_{1}}{2\alpha^{2}}-\frac{\alpha,_{12}}{\alpha
}=0.\label{aee5}
\end{gather}
In its turn the last system is equivalent to
\begin{gather*}
\mathbf{H}=\alpha\mathbf{M}_{\,}\\
f=\frac{c}{4}\triangle\alpha\\
\partial_{1}\left[  \ln\left|  f\right|  -\frac{1}{2}\left(  \ln\alpha
+\ln\frac{\left|  \nabla\left(  \alpha\right)  \right|  ^{2}}{\alpha^{2}
}\right)  \right]  =-\vartheta_{2}\\
\partial_{2}\left[  \ln\left|  f\right|  -\frac{1}{2}\left(  \ln\alpha
+\ln\frac{\left|  \nabla\left(  \alpha\right)  \right|  ^{2}}{\alpha^{2}
}\right)  \right]  =\vartheta_{1}
\end{gather*}
where $c=-\frac{1}{s^{2}m_{22}}$ and $\vartheta_{1}$ and $\vartheta_{2}$
are the partial derivatives of
\[
\vartheta=\arctan\frac{\alpha,_{2}}{\alpha,_{1}}.
\]
These equations show that $\vartheta$ and $\ln\frac{\sqrt{\alpha}\left|
f\right|  }{\left|  \nabla\left(  \alpha\right)  \right|  }$ are conjugated
harmonic functions so that the above system can be brought to the form:
\begin{gather}
\triangle\left(  \vartheta\right)  =0\nonumber\\
\frac{\alpha,_{2}}{\alpha,_{1}}=\tan\vartheta\label{r2}\\
\ln\frac{\sqrt{\alpha}\left|  \triangle\left(  \alpha\right)  \right|
}{\left|  \nabla\left(  \alpha\right)  \right|  }=\Phi\label{r3}\\
f=\frac{c}{4}\triangle\left(  \alpha\right) \label{r4}\\
\mathbf{H}=\alpha\mathbf{M}\nonumber
\end{gather}
where $\Phi$ is a harmonic function conjugated to $\vartheta$, that is a
primitive of the exact differential $1$-form $\omega=\vartheta_{1}
dx_{2}-\vartheta_{2}dx_{1}$. Now one can easily check that the above system
is  reduced to the \textit{tortoise} equation (see sec. \ref{se1.1})
\[
\beta+A\ln\left|  \beta-A\right|  =\Psi
\]
where $\beta^{2}=\alpha$, $\Psi$ is an arbitrary harmonic function and $A$
is an arbitrary constant. The functions $\vartheta$ and $\Phi$ are given,
respectively, by
\begin{align*}
\vartheta & =\arctan\frac{\Psi,_{2}}{\Psi,_{1}}\\
\Phi & =\ln\left|  \nabla\left(  \Psi\right)  \right|  .
\end{align*}
By summing up we give the components of the metric in terms of the adapted
holonomic frame $C=\left\{  dx_{1},dx_{2},dx,dy\right\}  $ with $x=x_{3},$
$y=x_{4},$ the $x_{\mu}$'s being the adapted coordinates introduced in
proposition \ref{prp5}.
\begin{proposition}
\label{gs3}Any $\mathcal{G}
_2$-integrable $4$-metric $g$ satisfying the vacuum
Einstein equations, and such that $\det{\bf F} >0$ and $h_{22}\neq0$, has
the following matrix form in the  adapted coordinates $\left(
x_{\mu}\right) $
\[
{\bf M}_{C}\left( g\right) =\left(
\begin{array}{cc}
\begin{array}{cc}
2f & 0 \\ 0 & 2f
\end{array}
& \mathbf{0} \\
\mathbf{0} & \beta^2 \left(
\begin{array}{cc}
s^{2}ky^{2}-2sly+m & -sky+l \\
-sky+l & k
\end{array}
\right)
\end{array}
\right)
\]
where
\begin{itemize}
\item$k$, $l$, $m$, are arbitrary constants such that $km-l^{2}=\pm
1$,  $k\neq0$,
\item
\begin{equation}
f=-\frac{1}{4s^{2}k}\triangle\left( \beta^2 \right) ,  \label{1fa3}
\end{equation}
\item$\beta$ is a solution of the tortoise equation
\begin{equation}
\beta+A\ln\left| \beta-A\right| =\Psi,  \label{2fa3}
\end{equation}
such that $\triangle\beta^2  \equiv
\left( \frac{\partial^{2}}{\partial x_{1}^{2}}+\frac{\partial^{2}}{
\partial x_{2}^{2}}\right) \beta^2 $ is everywhere
nonvanishing, $A$ and $\Psi$ being an arbitrary constant and an arbitrary
harmonic function.
\end{itemize}
\end{proposition}

\begin{remark}
Concerning the signature of $g$ and the character of the Killing fields, we
remark that:  If $\det\mathbf{M}=1$ (see Eq. (\ref{aee1})), then
$\mathbf{H}$ is either  positive or negative definite according to the sign
of $k$ as well as $g\left  ( Y,Y\right) $, and $g\left( X,X\right) $. Since
the sign of the constant $c$ is opposite to the one of $k$,  the signature
of $g$ is always equal to $0$.  If $\det\mathbf{M}=-1$, then $\mathbf{H}$
is  indefinite, $g\left( Y,Y\right) $ has the same sign as $k$ while the
sign of $g\left( X,X\right  ) $ varies with as a function of $y$. The
signature of $g$ is equal to $\pm  2$, so that these metrics are of
interest for General Relativity.
\end{remark}

Moreover, as in sec. \ref{se1.1} we have:
\begin{corollary}
The metric of the above proposition admits a third Killing field
\[
Z=\alpha e^{-sx_{2}}\left[ \left( m-sky\right)
\partial_{x}+\left( s^{2}ky^{2}-2smy+l\right) \partial_{y}\right],
\]
which together with $X$ and $Y$ generate a $3$-dimensional Lie algebra
isomorphic to $so\left(2,1\right)$
\[
\left[ X,Y\right] =sZ,\;\left[
X,Z\right] =-sZ,\;\left[ Y,Z\right] =-2skX.
\]
\end{corollary}

\subsubsection{ The case $h_{22}=0$.}

In this case the equations $R_{ia}=0$ are satisfied automatically while the
matrix $\mathbf{H}$ has the form
\[
\mathbf{H}=\left(
\begin{array}
[c]{cc}
\nu & \mu\\
\mu & 0
\end{array}
\right)  ,
\]
and $\alpha=\left|  \mu\right|  $. The remaining Einstein equations reduce
now to
\begin{gather}
\triangle\left(  \alpha\right)  =0\label{l01}\\
\left(  \alpha\partial_{1}w\right)  _{,1}+\left(  \alpha\partial_{2}w\right)
_{,2}=0\label{l02}\\
\triangle\left(  \ln\left|  f\right|  \right)  =\frac{1}{2}\left[  \left(
\frac{\alpha,_{1}}{\alpha}\right)  ^{2}+\left(  \frac{\alpha,_{2}}{\alpha
}\right)  ^{2}\right] \label{l03}\\
\alpha,_{1}\partial_{1}\left(  \ln\left|  f\right|  \right)  -\alpha
,_{2}\partial_{2}\left(  \ln\left|  f\right|  \right)  =\alpha,_{11}
-\alpha,_{22}-\frac{\alpha,_{1}^{2}-\alpha,_{2}^{2}}{2\alpha}\label{l04}\\
\alpha,_{2}\partial_{1}\left(  \ln\left|  f\right|  \right)  +\alpha
,_{1}\partial_{2}\left(  \ln\left|  f\right|  \right)  =2\alpha,_{12}
-\frac{\alpha,_{2}\alpha,_{1}}{\alpha}.\label{l05}
\end{gather}
where $\triangle=\frac{\partial^{2}}{\partial x_{1}^{2}}+\frac{\partial^{2}
}{\partial x_{2}^{2}}$ and $w=\frac{\nu}{\alpha}$. If $\alpha$ is a
solution  of Eq. (\ref{l01}), i.e., a harmonic function, then the general
solution of  Eq. (\ref{l03}) is
\[
f=\pm\alpha^{-\frac{1}{2}}e^{\psi}
\]
$\psi$ being a harmonic function. Substituting this expression in Eqs.
(\ref{l04}), (\ref{l05}) one gets
\begin{align*}
\alpha,_{1}\psi,_{1}-\alpha,_{2}\psi,_{2}  & =2\alpha,_{11}\\
\alpha,_{2}\psi,_{1}+\alpha,_{1}\psi,_{2}  & =2\alpha,_{12}
\end{align*}
the last relations are locally equivalent to
\[
\left|  \nabla\left(  \alpha\right)  \right|  ^{2}=ce^{\psi}
\]
$c$ being a constant. Therefore,
\begin{equation}
f=\pm\frac{\left|  \nabla\left(  \phi\right)  \right|  ^{2}}{\sqrt{\left|
D\phi+B\right|  }}\label{af4}
\end{equation}
where $\alpha=\left|  \mu\right|  =\left|  D\phi+B\right|  $, $A$ and $B$
are constants and $\phi$ a harmonic function such that $\alpha$ is
nonvanishing.  The equation (\ref{l02}) is a \textit{linear second order
partial differential  equation} and can be analyzed with standard methods.
  Thus, as the final result we have:
\begin{proposition}
\label{gs4}Any $\mathcal{G}_2
$-integrable $4$-metric $g$ satisfying the vacuum Einstein equations, and
such that $\det{\bf F}>0$ and $h_{22}=0$,  has  the following matrix form
in the  adapted coordinates $\left( x_1,x_2,x.y\right) $
\[
{\bf M}_{C}\left( g\right) =\left(
\begin{array}{cc}
\epsilon\frac{\left| \nabla\left( \phi\right) \right| ^{2}}{\sqrt{\left|
D\phi+B\right| }}\mathbf{1}_{2} & \mathbf{0} \\
\mathbf{0} & \left( D\phi+B\right) \left(
\begin{array}{cc}
-2sy+w & 1 \\
1 & 0
\end{array}
\right)
\end{array}
\right) .
\]
where $\varepsilon=\pm1$, $\phi$ is a harmonic function, $D$ and $B$ are
constants such that $\mu=D\phi+B$ is everywhere nonvanishing and $w$ is  a
solution of the equation
\[
\left( \mu w,_{1}\right) _{,1}+\left( \mu w,_{2}\right) _{,2}=0.
\]
\end{proposition}

In the considered case $\det\mathbf{H}$ is negative and the signature of
$g$ is equal to $\pm2$. The Killing vector field $Y$ is isotropic while the
sign  of $g\left(  X,X\right)  $ varies as a function of $y$. The Gauss
curvature  $K$ of the Killing leaves vanishes.
\begin{remark}
According to section \ref{kl}, an additional Killing field, say $Z$,
tangent to the Killing leaves and independent  of $X$ and $Y$, exists iff
$w$ is a  constant, say $w_0$. In  such a case it is given by
\begin{equation*}
Z=e^{-sx}\left[ -2 \partial_{x}+\left(
-2sy+w_0\right) \partial_{y}\right] ,
\end{equation*}
which generates together with $X=\partial_{x}$ and $Y=e^{sx}\partial_{y}$ a
3\textit{-dimensional Lie algebra} isomorphic to $\mathcal{K}il \left
(dx^2-dy^2\right)$:
\begin{equation*}
\left[ X,Y\right] =sY,\,\,\,\left[ X,Z\right] =-sZ,\,\,\left[ Y,Z\right]
=0.
\end{equation*}
\end{remark}

A canonical form for the equation (\ref{l02}) can be found by introducing
new coordinates, namely $\xi$ and $\eta$, by
\[
\xi=\alpha+\widetilde{\alpha},\hspace{0.2in}\eta=\alpha-\widetilde{\alpha}.
\]
in which Eq. (\ref{l02}) becomes
\[
\left(  \xi+\eta\right)  \left(  \frac{\partial^{2}}{\partial\xi^{2}}
+\frac{\partial^{2}}{\partial\eta^{2}}\right)  \left(  \widetilde{w}\right)
+\frac{\partial\widetilde{w}}{\partial\xi}+\frac{\partial\widetilde{w}
}{\partial\eta}=0,
\]
with $\widetilde{w}\left(  \xi,\eta\right)  \equiv w\left(  x_{1}\left(
\xi,\eta\right)  ,x_{2}\left(  \xi,\eta\right)  \right)  $, or,
alternatively,
\[
\left(  \frac{\partial^{2}}{\partial\xi^{2}}+\frac{\partial^{2}}{\partial
\eta^{2}}\right)  \left(  Z\right)  +\frac{1}{2\left(  \xi+\eta\right)  ^{2}
}Z=0.
\]
with
\[
Z=\sqrt{\xi+\eta}\widetilde{w}.
\]

For its geometrical meaning see \cite{SVV00}.
\section{The Abelian limit ($s=0$)\label{ABC}}

The solutions of the Einstein equations found in the previous section allow
one to get exact solutions of the Belinsky-Zahkarov case just by passing to
the ''Abelian limit'' $s=0$. Since the Abelian case was extensively studied
(see, for instance, \cite{BK70,Ge72,BZ78}) we shall limit ourself here
simply  to describe these solutions. In what follows we use the adapted
coordinates to  which the propositions refer and consider separately the
cases $h_{22}\neq0$  and $h_{22}=0$.
\paragraph{The case $h_{22}\neq0$}

With this assumption Eqs. (\ref{rai}) and, which is the same (\ref{rai2})
play the role of an ''ansatz'' when passing to the Abelian limit: So, in
that case  as in sec. \ref{se1.1} and sec. \ref{se1.2} one sees that
$H=\alpha M$, $M$  being a constant unimodular matrix.
\begin{itemize}
\item  If $\det\mathbf{F}<0$ then Eq. (\ref{na1}) becomes
\[
\alpha,_{12}=0
\]
and the remaining Einstein equations coincide with equations (\ref{h03})
and (\ref{h04}) as they appeared when analyzing the non-abelian situation
assuming  that $h_{22}=0$ and $\det F<0$ (see sec. \ref{se1.2}). Thus, the
same  procedure leads us to the following result:
\begin{equation}
\mathbf{M}_{C}\left(  g\right)  =\left(
\begin{array}
[c]{cc}
\begin{array}
[c]{cc} 2f & 0\\  0 & -2f
\end{array}
& \mathbf{0}\\
\mathbf{0} & \alpha\mathbf{M}
\end{array}
\right)  ,\label{ab1}
\end{equation}
where $\alpha$ and $f$ are given by
\begin{align}
\alpha=C_{1}F\left(  z_{1}+z_{2}\right)  +C_{2}G\left(  z_{1}-z_{2}\right)
+C\label{ab2-1}\\ f=\frac{F^{\prime}G^{\prime}}{\sqrt{\left|  \alpha\right|
}}\label{ab2-2}
\end{align}
$F$ and $G$ being arbitrary functions, $C$, $C_{1}$, $C_{2}$, arbitrary
constants such that $\alpha$ and $f$ are everywhere nonvanishing;
\item  If $\det\mathbf{F}>0$, then by referring to Eqs. (\ref{l01}
)-(\ref{l05}) one finds that
\[
\mathbf{M}_{C}\left(  g\right)  =\left(
\begin{array}
[c]{cc}
\epsilon\frac{\left|  \nabla\left(  \phi\right)  \right|  ^{2}}{\sqrt{\left|
D\phi+B\right|  }}\mathbf{1}_{2} & \mathbf{0}\\
\mathbf{0} & \left(  D\phi+B\right)  \mathbf{M}
\end{array}
\right)  .
\]
where $\epsilon=\pm1$, $\phi$ is a harmonic function, and $D$ and $B$ are
constants such that $D\phi+B$ is everywhere nonvanishing and $\mathbf{M}$
is  as above.
\end{itemize}

\paragraph{The case $h_{22}=0$.}

With this assumption the Abelian limit is, obviously, obtained from the
corresponding non-Abelian result (propositions \ref{gs2} and \ref{gs4})
just  by putting $s=0$. Namely:
\begin{itemize}
\item  If $\det\mathbf{F}<0$, then (proposition \ref{gs2})
\begin{equation}
\mathbf{M}_{C}\left(  g\right)  =\left(
\begin{array}
[c]{cc}
\begin{array}
[c]{cc}
-2f & 0\\
0 & 2f
\end{array}
& \mathbf{0}\\
\mathbf{0} & \mu\left(
\begin{array}
[c]{cc} w & 1\\  1 & 0
\end{array}
\right)
\end{array}
\right)  ,\label{ab3}
\end{equation}
where
\begin{align}
\mu=C_{1}F\left(  z_{1}+z_{2}\right)  +C_{2}G\left(  z_{1}-z_{2}\right)
+C\label{ab4-1}\\ f=\left|  \mu\right|
^{-\frac{1}{2}}F^{\prime}G^{\prime},\label{ab4-2}
\end{align}
$F$, $G$ and $C$, $C_{1}$, $C_{2}$, being arbitrary functions and
constants, respectively, such that $\mu$ and $f$ be everywhere nonvanishing
while $w$ is  an arbitrary solution of the equation
\[
\left(  \mu w,_{1}\right)  _{,2}+\left(  \mu w,_{2}\right)  _{,1}=0.
\]

\item  If $\det\mathbf{F}>0$, then (proposition \ref{gs4})
\begin{equation}
\mathbf{M}_{C}\left(  g\right)  =\left(
\begin{array}
[c]{cc}
\epsilon\frac{\left|  \nabla\left(  \phi\right)  \right|  ^{2}}{\sqrt{\left|
D\phi+B\right|  }}\mathbf{1}_{2} & \mathbf{0}\\
\mathbf{0} & \left(  D\phi+B\right)  \left(
\begin{array}
[c]{cc} w & 1\\  1 & 0
\end{array}
\right)
\end{array}
\right)  .\label{ab4}
\end{equation}
where $\epsilon=\pm1$, $\phi$ is a harmonic function, $D$ and $B$ are
arbitrary constants such that $\mu=D\phi+B$ is everywhere nonvanishing and
$w$  is an arbitrary solution of the equation
\[
\left(  \mu w,_{1}\right)  _{,1}+\left(  \mu w,_{2}\right)  _{,2}=0.
\]
\end{itemize}

\begin{remark}
It is worth to note that in the Abelian case the Gauss curvature of the
Killing leaves is equal to  zero.
\end{remark}

\section{Ricci-flat metrics admitting a $3$-dimensional Killing algebra with
bidimensional leaves\label{3G}}   Let $g$ be a metric and
$\mathcal{G}\,\,\,$be one of its Killing algebras. In  what follows, the
Killing algebra $\mathcal{G}$ will be called \textit{normal  }if the
restrictions of $g$ to its Killing leaves are non-degenerate.    Obviously,
a normal Killing algebra $\mathcal{G}$ is isomorphic to a  subalgebra of
$\mathcal{K}$\textit{il}$\left(  g|_{S}\right)  $ where $S$ is a  generic
Killing leaf of $\mathcal{G}$. Thus, when $\dim\mathcal{G}=3$ and the
Killing leaves are bidimensional, $\mathcal{G=K}il\left(  g|_{S}\right)  $.
As  it is easy to see, in this situation there are exactly five options for
$\mathcal{K}$\textit{il}$\left(  g|_{S}\right)  $ and, therefore, for
$\mathcal{G}$. Namely, they are:
\begin{equation}
so\left(  2,1\right)  ,\qquad\mathcal{K}\mathit{il}\left(  dx^{2}
-dy^{2}\right)  ,\qquad so\left(  3\right)  ,\qquad\mathcal{K}\mathit{il}
\left(  dx^{2}+dy^{2}\right)  ,\qquad\mathcal{A}_{3},\label{ls}
\end{equation}
where $\mathcal{A}_{3}$ is a $3$-dimensional Abelian Lie algebra. Since the
Lie algebra $\mathcal{A}_{3}$ belongs to the case treated in \cite{BZ78} it
will not be considered in the following.    Only two of these algebras,
namely $so\left(  2,1\right)  $ and $\mathcal{K}  $\textit{il}$\left(
dx^{2}-dy^{2}\right)  $, possess a non-commutative  bidimensional
subalgebra. Thus, one may expect that the corresponding Ricci  flat
$4$-metrics are among the solutions described in section
\textit{\ref{se1}}. It will be shown below that this is in fact true and that
they belong to one of the cases $h_{22}\neq0$, or $h_{22}=0$ with $w$ fixed
to be constant (see section \textit{\ref{se1}}).    As for the algebra
$\mathcal{K}$\textit{il}$\left(  dx^{2}+dy^{2}\right)  $,  it has only a
bidimensional commutative subalgebra and we shall see that the
corresponding Ricci-flat $4$-metrics are among the solutions described in
the  previous section \textit{\ref{ABC}} (the Abelian limit with
$h_{22}\neq0 $).    The following assertion generalizes lemma
\textit{\ref{lemma1}} (section
\textit{\ref{ikm}}).

\begin{lemma}
Let $X_{1}$, $X_{2}$ and $f_{1}X_{1}+f_{2}X_{2},\quad$ $f_{1},f_{2}$ $\in
C^{\infty}\left( M\right) $ be Killing fields of a metric $\left(
M,g\right) $. Then, supposing that $X_{1}$ and $X_{2}$ are independent,
either $f_{1}$ and $f_{2}$ are functionally independent, or $f_{1}$ and $
f_{2}$ are constant.
\end{lemma}

\begin{proof}
It results from relation (\ref{lfx}) taking into account $L_{X_{1}}\left(
g\right)  =$ $L_{X_{2}}\left(  g\right)  =0$ that
\begin{equation}
0=L_{f_{1}X_{1}+f_{2}X_{2}}\left(  g\right)  =i_{X_{1}}\left(  g\right)
df_{1}+i_{X_{2}}\left(  g\right)  df_{2}.\label{lg}
\end{equation}
Assuming, say, that $f_{2}=\varphi\left(  f_{1}\right)  $ we see that
\[
0=L_{f_{1}X_{1}+f_{2}X_{2}}\left(  g\right)  =\left(  i_{X_{1}}\left(
g\right)  +\varphi^{\prime}i_{X_{2}}\left(  g\right)  \right)  df_{1}
=i_{X_{1}+\varphi^{\prime}X_{2}}\left(  g\right)  df_{1},
\]
If $df_{1}\neq0$, then the last equality implies, obviously, $i_{X_{1}
+\varphi^{\prime}X_{2}}\left(  g\right)  =0$. In that case, $X_{1}
+\varphi^{\prime}X_{2}=0$ due to the non-degeneracy of $g$ in contradiction
with the assumed independence of $X_{1}$ and $X_{2}$. If on the contrary
$df_{1}=0$, then $df_{2}=0$ and the second alternative takes place.    Note
that it cannot happen that on a connected manifold $M$ the first
alternative takes place in $U_{1}\subset M$ and the second one in
$U_{2}\subset M$ if $\bigcap_{i}U_{i}\neq\emptyset$. It results from the
fact  that if a Killing field vanishes on an open subset of $M$, then it
vanishes everywhere.
\end{proof}

\begin{corollary}
\label{cor15}If $\overline{\cal{G}}$ is a $3$-dimensional Killing algebra
having bidimensional Killing leaves and the fields $X_{1}$, $X_{2}$ ,
$X_{3}$ generate it as a linear space, then almost everywhere $
X_{3}=f_{1}X_{1}+f_{2}X_{2}$ and $f_{1}$ and $f_{2}$ are functionally
independent.
\end{corollary}

\begin{proof}
The fields $X_{1}$ and $X_{2}$ are independent according to lemma
\ref{lemma1}. So they generate almost everywhere, say in $U$, the tangent
spaces to the Killing leaves. Thus, $X_{3}=f_{1}X_{1}+f_{2}X_{2},$\quad
$f_{i}\in$ $C^{\infty}\left(  U\right)  $. The possibility that $f_{1}$ and
$f_{2}$ are constant offered by lemma \textit{\ref{lemma1}} cannot occur in
this context since $X_{1}$, $X_{2}$ and $X_{3}$ are supposed to be linearly
independent.
\end{proof}

\begin{proposition}
\label{prp16}Any Killing algebra from the list (\ref{ls}
) having bidimensional Killing leaves is normal. Moreover, the distribution
$\mathcal {D}^{\bot}$ orthogonal to its Killing  leaves is integrable.
\end{proposition}

\begin{proof}
Below the notation of corollary \ref{cor15} is used. Since $df_{1}$and
$df_{2}$ are almost everywhere point-wise independent and one can deduce
easily from (\ref{lg}) that
\begin{equation}
i_{X_{1}}\left(  g\right)  =\lambda df_{2},\qquad i_{X_{2}}\left(  g\right)
=-\lambda df_{1},\label{ilg}
\end{equation}
being $g$ nondegenerate, $\lambda$ is almost everywhere non-vanishing. Let
now $Y$ be an almost everywhere different from zero vector field. Then the
equality
\[
i_{Y}\left(  i_{X_{1}}\left(  g\right)  df_{1}+i_{X_{2}}\left(  g\right)
df_{2}\right)  =0,
\]
which is an obvious consequence of (\ref{lg}), is equivalent to
\[
g\left(  X_{1},Y\right)  df_{1}+g\left(  X_{2},Y\right)  df_{2}=-Y\left(
f_{1}\right)  i_{X_{1}}\left(  g\right)  -Y\left(  f_{2}\right)  i_{X_{2}
}\left(  g\right)  .
\]
In view of (\ref{ilg}) it gives
\[
g\left(  X_{1},Y\right)  df_{1}+g\left(  X_{2},Y\right)  df_{2}=-\lambda
Y\left(  f_{1}\right)  df_{2}+\lambda Y\left(  f_{2}\right)  df_{1},
\]
so that
\[
g\left(  X_{1},Y\right)  =\lambda Y\left(  f_{2}\right)  ,\qquad g\left(
X_{2},Y\right)  =-\lambda Y\left(  f_{1}\right)  .
\]
Hence $Y\left(  f_{1}\right)  =Y\left(  f_{2}\right)  =0$ iff $g\left(
X_{1},Y\right)  =g\left(  X_{2},Y\right)  =0$, \textit{i.e.,} such fields
$Y$  are orthogonal to the Killing leaves and \textit{vice versa}. If $Y$
is  tangent to the Killing leaves, then
\[
Y\left(  f_{1}\right)  =Y\left(  f_{2}\right)  =0\Longleftrightarrow Y=0,
\]
since by the above corollary applied to the case $M=S$, $df_{i}|_{S}$ is
nondegenerate for a generic Killing leaf $S$. This proves that the fields
$Y$  such that $Y\left(  f_{1}\right)  =Y\left(  f_{2}\right)  =0$ are
transversal  to the Killing leaves and that $g|_{S}$ is non-degenerate for
a generic  Killing leaf $S$. Thus $\mathcal{G}$ is normal.    Finally note
that the distribution $\widetilde{\mathcal{D}}$ spanned by the  vector
fields $Y$ such that $Y\left(  f_{1}\right)  =Y\left(  f_{2}\right)
=0$ is of co-dimension \textit{2} since $df_{1}$ and $df_{2}$ are independent
almost everywhere. Being both transversal and orthogonal to the Killing
leaves, $\widetilde{\mathcal{D}}$ coincides with $\mathcal{D}^{\bot}$ by a
dimension argument.
\end{proof}

\begin{corollary}
The solutions found in section \ref{se1} exhaust all local Ricci-flat $4$
-metrics admitting a Killing algebra isomorphic to  $so\left( 2,1\right ) $
or to  $Kil\left( dx^{2}-dy^{2}\right) $.
\end{corollary}

\begin{proof}
As we already noticed, the first two algebras possess non-Abelian
bidimensional subalgebras and according to the previous proposition the
distribution $\mathcal{D}^{\bot}$ orthogonal to Killing leaves is
transversal  to them and integrable.
\end{proof}

\subsection{$\mathcal{K}$\textit{il}$\left(  dx^{2}+dy^{2}\right)  $-invariant
Ricci-flat metrics}   As it has been already noticed, the algebra
$\mathcal{K}$\textit{il}$\left(  dx^{2}+dy^{2}\right)  $ has a
bidimensional commutative subalgebra. We shall  see that the corresponding
Ricci-flat $4$ -metrics are among the solutions of  previous section
\textit{\ref{ABC}} (the Abelian limit with $h_{22}\neq0$).    First, let
$\mathcal{G}$ be a Killing algebra isomorphic to $\mathcal{K}
$\textit{il}$\left(  dx^{2}+dy^{2}\right)  $ and let
$X_{i},$\quad$i=1,2,3$,  be its standard basis, \textit{i.e.,}
\[
\left[  X_{1},X_{2}\right]  =0,\quad\left[  X_{1},X_{3}\right]  =X_{2}
,\quad\left[  X_{2},X_{3}\right]  =-X_{1}.
\]
With the notation of corollary \ref{cor15}, let
$X_{3}=f_{1}X_{1}+f_{2}X_{2}$. Then
\begin{align*}
X_{2}  & =\left[  X_{1},X_{3}\right] \\ & =\left[
X_{1},f_{1}X_{1}+f_{2}X_{2}\right] \\  & =X_{1}\left(  f_{1}\right)
X_{1}+X_{1}\left(  f_{2}\right)  X_{2}
\end{align*}
and
\begin{align*}
X_{1}  & =\left[  X_{3},X_{2}\right] \\ & =\left[
f_{1}X_{1}+f_{2}X_{2},X_{2}\right] \\  & =-X_{2}\left(  f_{1}\right)
X_{1}-X_{2}\left(  f_{2}\right)  X_{2},
\end{align*}
so that, for the independence (section \ref{ikm}, lemma \ref{lemma1}) of
$X_{1}$ and $X_{2}$, implies that we have
\begin{align*}
X_{1}\left(  f_{1}\right)   & =0,\quad X_{1}\left(  f_{2}\right)  =1\\
X_{2}\left(  f_{1}\right)   & =-1,\quad X_{2}\left(  f_{2}\right)  =0.
\end{align*}
Joining to $f_{1}$, $f_{2}$ a couple of independent functions $z_{1}$,
$z_{2} $ such that $X_{i}\left(  z_{j}\right)  =0,\forall i,j$, one gets a
local  chart on $M$. Taking into account the above relations and passing to
the  standard coordinate notation $x=f_{1}$, $y=f_{2}$, we see that in the
chart  $\left(  x,y,z_{1},z_{2}\right)  $
\[
X_{1}=\partial_{y},\quad X_{2}=-\partial_{x},\quad X_{3}=x\partial
_{y}-y\partial_{x}.
\]

Introducing on $S$ \textit{polar coordinates }$\left(  r,\varphi\right)  $,

\textit{i.e.}, $x=r\cos\varphi,\,\,\,y=r\sin\varphi$, the above fields read
as
\[
X_{1}=\sin\varphi\partial_{r}+\frac{\cos\varphi}{r}\partial_{\varphi},\quad
X_{2}=\cos\varphi\partial_{r}+\frac{\sin\varphi}{r}\partial_{\varphi},\quad
X_{3}=\partial_{\varphi}.
\]
Then, in view of proposition \textit{\ref{prp16}}, a direct computation
similar to the one of section \textit{\ref{ikm}} shows that any
$\mathcal{G}  $-invariant metric has in the adapted local chart $\left(
z_{1}  ,z_{2},r,\varphi\right)  $ the form
\[
g=2f\left(  dz_{1}^{2}+\varepsilon dz_{2}^{2}\right)  +\mu\left(  z_{1}
,z_{2}\right)  \left[  dr^{2}+r^{2}d\varphi^{2}\right]  ,
\]
and, therefore, belongs to the class of metrics considered in section
\textit{\ref{ABC}} with definite $\mathbf{H}$ and $h_{22}\neq0$.

Thus, we have:
\begin{corollary}
The solutions found in section \ref{ABC} exhaust all local Ricci-flat
$4$-metrics  admitting a Killing algebra isomorphic to
$Kil\left(dx^{2}+dy^{2}\right) $.
\end{corollary}

\subsection{$so\left(  3\right)  $-invariant Ricci-flat metrics}

The above results lead to expect that Ricci-flat $4$-metrics admitting a
Killing algebra isomorphic to $so\left(  3\right)  $ with $2$-dimensional
leaves can be described essentially in the same way as it was done in
section
\textit{\ref{se1}} with respect to those admitting a Killing algebra
isomorphic to $so\left(  2,1\right)  $. The details are as follows   First,
let $\mathcal{G}$ be a Killing algebra isomorphic to $so\left(  3\right)  $
and let $X_{i},\,\,i=1,2,3$, be its standard basis, \textit{i.e.}
\[
\left[  X_{1},X_{2}\right]  =X_{3},\quad\left[  X_{2},X_{3}\right]
=X_{1},\quad\left[  X_{3},X_{1}\right]  =X_{2}.
\]
In the notation of corollary \ref{cor15} let $X_{3}=f_{1}X_{1}+f_{2}X_{2}$.
Then
\begin{align*}
X_{1}  & =\left[  X_{2},X_{3}\right]  =\left[  X_{2},f_{1}X_{1}+f_{2}
X_{2}\right] \\  & =X_{2}\left(  f_{1}\right)  X_{1}+f_{1}\left[
X_{2},X_{1}\right]  +X_{2}\left(  f_{2}\right)  X_{2}\\  & =\left(
X_{2}\left(  f_{1}\right)  -f_{1}^{2}\right)  X_{1}+\left(  X_{2}\left(
f_{2}\right)  -f_{1}f_{2}\right)  X_{2}.
\end{align*}
Since $X_{1}$ and $X_{2}$ are independent (lemma \ref{lemma1})
\begin{equation}
X_{2}\left(  f_{1}\right)  -f_{1}^{2}=1,\quad X_{2}\left(  f_{2}\right)
-f_{1}f_{2}=0.\label{R1}
\end{equation}
Similarly, from the relation $\left[  X_{3},X_{1}\right]  =X_{2}$ one finds

\begin{equation}
X_{1}\left(  f_{1}\right)  +f_{1}f_{2}=0,\quad X_{1}\left(  f_{2}\right)
+f_{2}^{2}=-1.\label{R2}
\end{equation}
Joining to $f_{1}$, $f_{2}$ a couple of independent functions $z_{1}$,
$z_{2} $ such that $X_{i}\left(  z_{j}\right)  =0,\forall i,j$, one gets a
local  chart on $M$. Taking into account relations (\ref{R1}) and
(\ref{R2}) and  passing to the standard coordinate notation $x=f_{1}$,
$y=f_{2}$, we see that  in the chart $\left(  x,y,z_{1},z_{2}\right)  $
\[
X_{1}=-xy\partial_{x}-\left(  1+y^{2}\right)  \partial_{y},\quad
X_{2}=\left( x^{2}+1\right)  \partial_{x}+xy\partial_{y},\quad
X_{3}=y\partial
_{x}-x\partial_{y}.
\]

In\textit{\ }the \textit{geographic coordinates }$\left(  r,\varphi\right)
$ , \textit{i.e.},
$x=\tan\vartheta\cos\varphi,\,\,\,y=\tan\vartheta\sin\varphi  $, the above
fields read as
\[
X_{1}=-\frac{\cos\varphi}{\tan\vartheta}\partial_{\varphi}-\sin\varphi
\partial_{\vartheta},\quad X_{2}=-\frac{\sin\varphi}{\tan\vartheta}
\partial_{\varphi}+\cos\varphi\partial_{\vartheta},\quad X_{3}=-\partial
_{\varphi}.
\]
Then, in view of proposition \textit{\ref{prp16}}, a direct computation
similar to the one of section \textit{\ref{ikm}} shows that any
$\mathcal{G}  $-invariant metric has in the adapted local chart $\left(
z_{1}  ,z_{2},\vartheta,\varphi\right)  $ the form
\begin{equation}
g=f\left(  dz_{1}^{2}+\varepsilon dz_{2}^{2}\right)  +\alpha\left(
z_{1},z_{2}\right)  \left[  d\vartheta^{2}+\sin^{2}\vartheta d\varphi
^{2}\right] \label{sch1}
\end{equation}

The Ricci tensor of the above metric can be easily computed as in section
\textit{\ref{ctr}} and the corresponding Einstein equations lead to the same
equations for $f$ and $\alpha$ $\equiv r^{2}$ as already found in section
\textit{\ref{se1}} in the case $h_{22}\neq0$. Namely,
\begin{equation}
f=-\frac{1}{2}\left(  \frac{\partial^{2}}{\partial z_{1}^{2}}+\varepsilon
\frac{\partial^{2}}{\partial z_{2}^{2}}\right)  \left(  r^{2}\right)
,\label{sch3}
\end{equation}
\begin{equation}
r+A\ln\left|  r-A\right|  =u,\label{sch2}
\end{equation}
with $\varepsilon=\pm1$, $A$ being an arbitrary constant and $u$ being an
arbitrary function satisfying the equation
\[
\left(  \frac{\partial^{2}}{\partial z_{1}^{2}}+\varepsilon\frac{\partial^{2}
}{\partial z_{2}^{2}}\right)  \left(  u\right)  =0.
\]
Additionally, $f$ is required to be nonvanishing.
\begin{remark}
In the case $\varepsilon=-1$,these solutions are locally diffeomorphic to
the Schwarzschild solution. This will be discussed in \cite{SVV00}.
\end{remark}

Below, the graph of the left hand side of the equation (\ref{sch2}) is
reported for the values $A=2$ and $A=-2$.
\[
\begin{array}
[c]{cc}
\includegraphics[natheight=2.000300in,natwidth=2.332400in,
height=2.0003in,width=2.3324in] {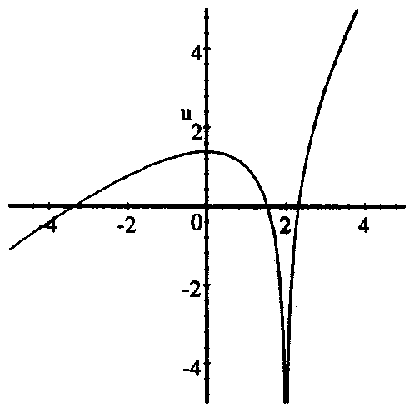} &
{\includegraphics[natheight=2.000300in,natwidth=2.332400in,
height=2.0003in,width=2.3324in]{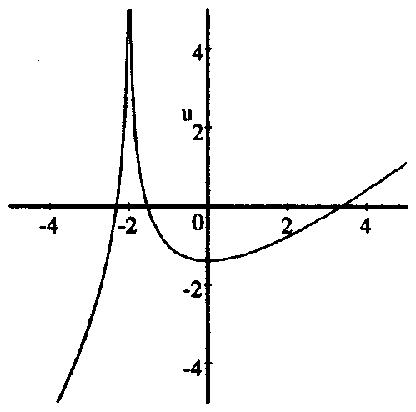}}
\\
u=r+2\ln\left|  r-2\right|  & u=r-2\ln\left|  r+2\right|
\end{array}
\]

One can see that for $A\neq0$ there exactly three possibilities for
$r=r\left(  u\right)  $ that correspond to the intervals of monotonicity of
$u\left(  r\right)  $. For instance, for $A>0$ these are $\left]
-\infty,0\right[  $, $\left]  0,A\right[  $, and $\left]  A,\infty\right[  $.
In these regions the corresponding metric (\ref{sch1}) is regular and has
some singularities along the curves $r=0$ and $r-A=0$.    Some geometrical
peculiarities of the obtained \textit{local} solutions show  how to match
them together in order to get \textit{global nonextendible  }Einstein
metrics. To this purpose, in \cite{SVV00} a formalism is developed  which
allows to construct, starting from known solutions, ''new'' global ones and
to describe their singularities as well. For instance, by extracting the
\textit{square root \ }of the Schwarzschild solution, one easily finds an
Einstein metric which describes \textit{parallel universes. }Other\textit{\
} examples which illustrate some aspects of our approach can be found in
\cite{SVV00}. We stress that it generalizes naturally to some other
situations  as, for instance, \textit{cosmological Einstein metrics
}satisfying  assumptions I and II (work in progress).
\textbf{Acknowledgments}

Two of the authors (G.S. and G.V.) wish to thank G. Bimonte, B. Dubrovin
and G.Marmo for interesting discussions.


\begin{thebibliography}{99}
\bibitem{AL92}B.N. Aliev and A.N.Leznov, \textit{Exact solutions of the vacuum
Einstein's equations,} J. Math. Phys., vol. 33, n. 7 (1992)2567-2573
\bibitem {Be91}V.A. Belinsky, \textit{Gravitational breather and topological
properties of gravisolitons,} Physical Review D, vol. 44 \ n.10
(1991)3109-3115
\bibitem {BK70}V.A.Belinsky and I.M.Khalatnikov, \textit{General solution of
the gravitational equations with a physical singularities,} Sov. Phys. Jetp
vol. 30, 6 (1970)
\bibitem {BZ78}V.A.Belinsky and V.E.Zakharov, \textit{Integration of the
Einstein equations by means of the inverse scattering problem technique and
construction of exact soliton solutions, }Sov. Phys. Jetp vol. 48, 6 (1978)

\bibitem {BZ79}V.A.Belinsky and V.E.Zakharov, \textit{\ Stationary
gravitational solitons with axial symmetry\ }Sov. Phys. Jetp vol. 50, 1
(1979)
\bibitem {Ch97}F. J. Chinea, \textit{New first integral for twisting type-N
vacuum gravitational fields with two non-commuting Killing vectors,} Class.
Quantum Grav. vol. 15 (1998)367-371
\bibitem {Ge72}R. Geroch, \textit{A method for generating new solutions of
Einstein's equation. II, }J.Math.Phys. vol. 13, 3 (1972).
\bibitem {Ha88}M. Hallisoy, \textit{Studies in space-times admitting two
spacelike Killing vectors, }J.Math.Phys. vol. 29, 2 (1988).
\bibitem {Pe69}A. Z. Petrov, \textit{Einstein spaces, (}Pergamon
Press,\textit{\ }New York 1969)   D. Kramer, H. Stephani, E. Herlt, M.
MacCallum, \textit{Exact  solutions of Einstein field equations,} Cambridge
University Press  1980.
\bibitem {SVV00}G.\ Sparano, G. Vilasi and A.\ M.\ Vinogradov, \textit{Gravitational
fields with a non-Abelian, bidimensional Lie algebras of symmetries}, Phys.
Lett B 513, (2001)142-146, \textit{Vacuum  Einstein metrics with
bidimensional Killing leaves},
\textit{II-Global aspects}, Differential Geometry and its
applications, to appear.
\end{thebibliography}
\end{document}